# Kinetic-Ion Simulations Addressing Whether Ion Trapping Inflates Stimulated Brillouin Backscattering Reflectivities

B.I. Cohen and E. A. Williams

University of California Lawrence Livermore National Laboratory

P.O. Box 808, Livermore, CA 94551

and H. X. Vu

University of California, San Diego, La Jolla, CA 92093

Abstract

An investigation of the possible inflation of stimulated Brillouin backscattering (SBS) due to ion kinetic effects is presented using electromagnetic particle simulations and integrations of three-wave coupled-mode equations with linear and nonlinear models of the nonlinear ion physics. Electrostatic simulations of linear ion Landau damping in an ion acoustic wave, nonlinear reduction of damping due to ion trapping, and nonlinear frequency shifts due to ion trapping establish a baseline for modeling the electromagnetic SBS simulations. Systematic scans of the laser intensity have been undertaken with both one-dimensional particle simulations and coupled-mode-equations integrations, and two values of the electron-to-ion temperature ratio (to vary the linear ion Landau damping) are considered. Three of the four intensity scans have evidence of SBS inflation as determined by observing more reflectivity in the particle simulations than in the corresponding three-wave mode-coupling integrations with a linear ion-wave model, and the particle simulations show evidence of ion trapping.

# I. INTRODUCTION

The nonlinear interaction of intense, coherent electromagnetic waves in high temperature plasmas plays an important role in laser fusion.[1,2] Stimulated backscattering of the incident electromagnetic wave by electron plasma waves, stimulated Raman backscattering (SRS), and by ion acoustic waves, stimulated Brillouin backscattering (SBS), are of particular interest because the backscattering can damage the final optics of the laser and degrade the symmetric illumination (direct or indirect) of the laser-fusion target.[3,4] Stimulated backscattering instabilities in laser plasmas have been the object of more than thirty years of intense research in experiments, analytical theory, and simulations. The new work presented here addresses specific aspects of the saturation of stimulated Brillouin backscattering in which nonlinear kinetic ion physics is important.

Our previous research on SBS has examined the role of nonlinear ion physics in the saturation of SBS in one and two spatial dimensions, and with the inclusion of ion-ion collisions and spatial inhomogeneity.[5,6,7,8,9] Vu and co-workers have elucidated the effects of wave breaking and ion trapping in SBS with particle simulations.[10] Recent work by Vu, Dubois, and Bezzerides has demonstrated with particle simulations and detailed modeling that SRS can exhibit reflectivities significantly in excess of those derived from models in which the electron plasma wave is assumed to be small amplitude and nonlinearities are absent.[11] The research by Vu and co-workers on SRS was motivated by experimental observations of SRS reflectivities much in excess of linear theory reported by Montgomery and co-workers.[12] The simulations and modeling in Ref. 11 make the case that electron trapping, which nonlinearly reduces electron Landau damping[13] and produces a nonlinear frequency shift[14,15,16] in the electron plasma wave, can produce kinetic inflation of SRS because the linear SRS gain depends inversely on the damping of the electron plasma wave when the electron plasma wave is strongly damped. The effects of ion trapping on ion acoustic waves are quite analogous to electron trapping effects on electron plasma waves, [17,5] Evidence of ion trapping has been observed frequently in simulations[4-10] of SBS and in experiments.[18] We are thus motivated to examine whether ion trapping in the SBS ion acoustic waves can nonlinearly reduce ion wave damping and inflate the SBS reflectivities.

Here we present an investigation of SBS using electromagnetic particle simulations including nonlinear ion kinetic effects and integrations of three-wave coupled-mode equations with a reduced model of the nonlinear ion physics addressing whether kinetic inflation of SBS can occur. To establish a baseline for modeling the electromagnetic SBS simulations, we have undertaken electrostatic simulations to directly measure the linear ion Landau damping of small-amplitude ion acoustic waves and the nonlinear reduction of damping and emergence of a nonlinear frequency shift due ion trapping in large-amplitude ion waves. Systematic scans of the laser intensity have been undertaken with both one-dimensional electromagnetic particle simulations and a coupled-mode-equations model to study SBS; and two values of the electron-to-ion temperature ratio (to vary the linear ion Landau damping) are considered. Three of the four intensity scans yield significant evidence of SBS inflation as determined by observing more reflectivity in the particle simulations (which show evidence of ion trapping) than in the corresponding three-wave mode-coupling integrations with a linear (small-amplitude) ion-wave model. Integrations of the three-wave mode-coupling equations with a simplified nonlinear model incorporating ion-trapping effects are useful in elucidating the effects of nonlinearities on SBS. Our studies also demonstrate the importance of kinetic simulations of SBS in giving guidance to reduced models.

The paper is organized as follows. Section I introduces the subject matter and its motivation. Section II comments on how collisions can affect ion-trapping effects and describes the particle simulation and three-wave-coupling models. Section III presents one-dimensional electrostatic simulations of small and large-amplitude ion waves to establish the linear damping of the small-amplitude waves and the nonlinear damping and frequency shifts of large-amplitude waves. In Sec. IV we report the results of four scans of laser intensity in electromagnetic particle simulations of SBS and the corresponding three-wave mode-coupling modeling. Conclusions are presented in Sec. V.

## II. PARTICLE SIMULATION AND THREE-WAVE MODE COUPLING MODEL
### A. Particle-in-cell simulations with ion-ion collisions

Particle-in-cell simulations with kinetic ions and Boltzmann electrons described elsewhere[5,11] are used here to study stimulated Brillouin backscattering (SBS) with nonlinear kinetic ion effects. The electrons are modeled as a fluid with a Boltzmann response to the longitudinal electric fields. The fluid electrons respond to the perpendicularly polarized pump and scattered electromagnetic waves, provide the transverse current in Maxwell's equations, and produce the ponderomotive potential that perturbs the electron density and drives the SBS ion waves. There is no collisional absorption of the electromagnetic waves in this model. We impose an explicit Fokker-Planck ion-ion collision operator[9] in some of the simulations. An important consequence of ion-ion collisions is that they provide a detrapping mechanism for resonant ions that are trapped in the electric potential troughs of the ion acoustic waves. When $\omega_b \tau_{coll} > 1$, where $\omega_b$ is the trapped ion bounce frequency and $\tau_{coll}$ is the collisional detrapping time, nonlinear effects due to trapping should be effective; and when the opposite inequality holds, the characteristic collision time $\tau_{coll}$ is too short to allow the ions to be trapped.

Two estimates of $\tau_{coll}$ can be obtained by considering the separate effects of parallel collisional diffusion and perpendicular scattering.[19,20] Divol, et al.[19] have estimated the characteristic collision time based on the time required for an ion to parallel diffuse out of the trapping region in the ion velocity distribution. From the condition $\omega_b \tau_{coll} > 1$, we estimate a relative electron density perturbation amplitude $\delta n/n$ for the ion trapping to dominate over parallel diffusion:

$$\nu_\parallel = \nu_0 (v_{thi}^2/c_s^2), \nu_0 = 4\pi Z^4 e^4 n_i \Lambda_{ii}/m_i^2 c_s^3, \omega_b = \omega_s (\delta n/n)^{1/2}$$
$$d\ln\delta^2/dt = -\nu_\parallel(c_s^2/\delta^2) = -\nu_{eff} \equiv -\tau_{coll}^{-1}, \delta \sim v_{trap} \sim c_s(\delta n/n)^{1/2} \quad (1)$$
$$\Rightarrow \delta n/n > \left(\frac{\nu_0^{NRL}}{\omega_s}\right)^{2/3} \frac{1}{(ZT_e/T_i)^{5/3}}$$

where $\nu_\parallel$ is the parallel diffusion rate, $c_s$ is the ion sound speed, $v_{thi}$ is the ion thermal velocity, $Z$ is the ion charge state, $m_i$ is the ion mass, $\delta$ is the ion trapping (plateau) width, $e$ is the electron charge, $\Lambda_{ii}$ is the Coulomb logarithm, $\nu_0^{NRL} = 4\pi Z^4 e^4 n_i \Lambda_{ii}/m_i^2 v_{thi}^2$, $\omega_s$ is the ion acoustic frequency (later in the discussion we will include dispersive corrections to the ion acoustic frequency), $n_i$ is the ion charge density, $T_{e,i}$ are the electron and ion temperatures, $ZT_e/T_i = c_s^2/v_{thi}^2$, $\omega_b$ is the ion trapping frequency, and $v_{trap}$ is the ion trapping velocity.

In Ref. 20 there is an estimate of the time needed for perpendicular ion scattering to undo the distortion due to trapping and restore the ion velocity distribution to Maxwellian.[20] From this estimate, we deduce a second condition on the ion wave $\delta n/n$ such that $\omega_{trap}\tau_{coll} > 1$:

$$\nu_\perp^i = \frac{2\sqrt{\pi}}{(ZT_e/T_i)^{3/2}} \nu_0^{NRL}, \tau_{coll} \approx \frac{2}{\nu_\perp^i}\frac{\delta}{c_s} \approx \frac{4}{\nu_\perp^i}\left(\frac{\delta n}{n}\right)^{1/2} \Rightarrow \delta n/n > O(1) \frac{\nu_0^{NRL}}{\omega_s} \frac{1}{(ZT_e/T_i)^{3/2}} \quad (2)$$

By comparing the two estimates for $\tau_{coll}$ given in Eqs.(1) and (2), we deduce that parallel diffusion is more effective than perpendicular scattering as a detrapping mechanism when

$$\tau_{coll,\parallel}/\tau_{coll,\perp} = (ZT_e/T_i)(\delta n/n)^{1/2} < 1 \quad (3)$$

In our nominally "collisionless" particle simulations, there are collisions due to the discrete-particle representation of the plasma.[21] A detailed analysis of these effects is given in Ref. 11 in the context of simulations of stimulated Raman backscattering in which electron trapping occurs. The arguments in Ref. 11 can be applied to our simulations of SBS with kinetic ions. An effective collision frequency $\nu_{ii}^{eff}$ for ions resonant with the ion wave can be estimated from a diffusion rate due to incoherent thermal fluctuations; in one spatial dimension these arguments yield:

$$\nu_{ii}^{eff}/\omega_{pi} \approx D_\parallel/\omega_{pi}\nu_{thi}^2 \approx O(1) \frac{1}{2n_i\lambda_i} \frac{\nu_{thi}}{\nu} \to O(1) \frac{1}{2n_i\lambda_i}\frac{\nu_{thi}}{c_s} \to \nu_{ii}^{eff}/\omega_s \approx O(1) \frac{1}{n_i\lambda_e} < 10^{-2} \quad (4)$$

where $\lambda_{i,e}$ are the ion and electron Debye lengths, $n_i$ is the ion superparticle density, and Eq.(4) has been evaluated for our simulations in which $k\lambda_e \sim 0.4$ and $n_i\lambda_e = 256$.

**B. Linear and nonlinear models of stimulated Brillouin backscattering**

In a uniform plasma slab for ion waves whose damping dominates convection, (no absolute instability, $\gamma_0 < \gamma_s v_g/2\sqrt{c_s v_g}$ in the absence of light-wave damping, where $\gamma_0^2 = k_0^2 v_0^2 \omega_{pi}^2/8k_0 c_s \omega_0$ is the square of the uniform-medium temporal growth rate for SBS[1-3]), the intensity gain exponent for convective amplification of the SBS backscattered electromagnetic wave is given by[1-3]

$$G_{SBS}^I = \frac{1}{8}\frac{v_0^2}{v_e^2}\frac{n_e}{n_c}\frac{\omega_s}{\gamma_s}\frac{\omega_0 L_x}{v_g(1+k^2\lambda_e^2)} \qquad (5)$$

The gain exponent is an important parameter in characterizing the conditions for SBS. Here $\gamma_s$ is the ion wave damping rate, $L_x$ is the length of the plasma, $\omega_0$ is the laser frequency, $v_0$ is the electron quiver velocity in the laser field, $v_e$ is the electron thermal velocity, $v_g$ is the group velocity of the backscattered wave, $\omega_{pi}$ is the ion plasma frequency, $k_0$ is the wavenumber of the laser, and $n_e/n_c$ is the ratio of the electron density to the critical density (where the laser frequency equals the electron plasma frequency).

If the ion wave in SBS remains relatively small in amplitude and is heavily damped ($\gamma_0 < \gamma_s v_g/2\sqrt{c_s v_g}$), then Eq.(5) describes the exponential amplification of the backscattered wave for weak backscatter, *i.e.*, power reflectivities $R<<1$. The backscatter amplification comes at the expense of pump depletion, which must be taken into consideration for finite $R$. An analytical solution for pump depletion in one spatial dimension and in the limit $\gamma_0 < \gamma_s v_g/2\sqrt{c_s v_g}$ has been given by Tang[22] describing the reflectivity if a steady state is reached:

$$R(1-R)+R_0R=R_0\exp[(1-R)G] \qquad (6)$$

where $G$ for SBS is the gain given by Eq.(5) and $R_0$ is the ratio of the intensity of the backscattered wave to the incident laser intensity at the boundary opposite the incident plane. For $R<<1$, $R\sim R_0\exp(G)$. Equation (6) is a transcendental equation for $R$ given $R_0$ and $G$, which is most easily solved for $G$ as a function of $R$ given $R_0$. We shall compare the Tang formula, Eq.(6), to the time-averaged reflectivities of the saturated SBS in our simulations. We note that the only nonlinearity in the Tang formula is pump depletion, and the ion wave is taken to be linear.

If ion trapping is significant, a number of new effects emerge that change the character of the ion wave. The ion wave Landau damping is expected to be reduced as the trapping distorts the velocity distribution, and there is also a nonlinear frequency shift,[5-9,13-17,19] $\Delta\omega_{nl}/\omega_s = -\eta|\delta n/n|^{1/2}$, where $\eta$ is a dimensionless coefficient that in the simplest limit is given by[7,14-17] $\eta = -O(1)v_\phi^3 \partial^2 f/\partial v^2$ where $f$ is the unperturbed velocity distribution function. While the reduction in the damping might be expected to increase the SBS gain, the frequency shift may produce a gain reduction due to detuning, *i.e.,* intuitively, one might expect $G \propto 1/\gamma_s \Rightarrow 1/(\gamma_s^2 + \Delta\omega_{nl}^2)^{1/2}$. If there is little or no detuning effect, and if there is no new ion damping mechanism engendered by other nonlinearities, then there is a basis for expecting an increase in the effective SBS gain due to the effects of trapping through the reduction in ion wave Landau damping. In this situation an inflation of the SBS reflectivity might be observed as in the case of SRS reported experimentally[12] and in recent simulation work.[11] Understanding the linear and nonlinear ion wave dissipation is a key element in assessing whether inflation of SBS is occurring.

Particle simulations of SBS typically see bursty reflectivities and no steady state. The ion-wave model in the particle simulations naturally retains convection, whose neglect as in the Tang model would be especially suspect if trapping is causing a reduction of the ion wave damping. Thus, we are motivated to model SBS with a set of one-dimensional coupled-mode equations retaining time dependence and convection, and with some nonlinear modifications in the ion-wave density perturbation equation to capture the effects of ion trapping. The coupled-mode equations derive straightforwardly from well-established theory:[1,3,4,23,24]

$$\left(\frac{\partial}{\partial t} + v_{g0}\frac{\partial}{\partial x}\right)a_0 = -ic_0(\delta n_e/n_e)a_1$$
$$\left(\frac{\partial}{\partial t} - v_{g1}\frac{\partial}{\partial x}\right)a_1^* = ic_1(\delta n_e/n_e)a_0^* \quad (7)$$
$$c_0 = \omega_{pe}^2/\omega_1, c_1 = \omega_{pe}^2/\omega_0, \omega_0 = \omega_1 + \omega_s, v_{g0,1} = k_{0,1}c^2/\omega_{0,1} \approx v_g, a_{0,1} = E_{0,1}$$

for the pump and backscattered electromagnetic waves, where $\omega_{pe}$ is the electron plasma frequency; $E_{0,1}$ are the slowly varying, complex-valued amplitudes for the transverse electric fields, $\delta n_e$ is the slowly varying, complex-valued amplitude of the electron density perturbation; and $v_g$ is the group velocity of the light waves; and equations for linear ion waves:

$$\left(\frac{\partial}{\partial t} + c_s\frac{\partial}{\partial x} + \gamma_s\right)\delta n_e/n_e = -ic_2 a_0 a_1^* + S_{noise}$$
$$c_2 = \frac{v_0^2 \omega_0 \omega_s}{4 v_e^2 E_{00}^2 \omega_1}, \gamma_s = \gamma_{LD} + v_{ii}, S_{noise} = noise \quad (8)$$

and nonlinear ion waves:

$$\left(\frac{\partial}{\partial t} + c_s\frac{\partial}{\partial x} - i\Delta\omega_{nl} + \gamma_{nl}\right)\delta n_e/n_e = -ic_2 a_0 a_1^* + S_{noise}$$
$$\gamma_{nl} = \gamma_{LD}\exp(-\int_0^t dt\frac{\omega_b}{2\pi}) + v_{ii}, \omega_b = \omega_s\left|\hat{\delta n_e}/n_e\right|^{1/2}, \hat{\delta n_e} = (\delta n_e - \delta n_{noise}, 0)_> \quad (9)$$
$$\Delta\omega_{nl} = -\eta\omega_b[1 - \exp(-\int_0^t dt\frac{\omega_b}{2\pi})]$$

where $\gamma_{nl}$ is the nonlinear ion wave damping rate, $\gamma_s$ is the linear ion wave damping, $\omega_b$ is the ion trapping frequency, $v_{ii}$ is the ion wave damping due to collisions, $S_{noise}$ represents the thermal noise source present in the particle simulations,[10,21] $\Delta\omega_{nl}$ is the nonlinear frequency shift, and $\Delta\omega_{nl}=0$ is an option in the nonlinear ion model to emphasize "inflation" effects. Based on the calculations in Refs. 13-16, the Landau damping nonlinearly relaxes over a few trapped ion bounce periods and the nonlinear frequency shift is established on the same time scale. The noise source is given by the simple model: $S_{noise}=A_{noise}\exp(i\Theta)$ with $\Theta=2\pi r(t)$, where $r(t)\in[0,1]$ is randomly chosen at every time step.

We integrate these equations with an explicit, finite-difference, predictor-corrector scheme using $\Delta x = v_g \Delta t$. The incident pump and backscattered waves in Eq.(7) are advected from left to right and right to left, respectively, along the characteristics of the equations with the right sides evaluated explicitly using a predictor-corrector iteration to approximately center the evaluation. The sound wave convection term is generally a small term for $\gamma_0 < \gamma_s v_g / 2\sqrt{c_s v_g}$ and is treated explicitly with central differencing in space. The high-frequency coupled-mode equations admit an action conservation law:

$$\frac{\partial}{\partial t}(J_0 + J_1) + (v_{g0}\frac{\partial}{\partial x}J_0 - v_{g1}\frac{\partial}{\partial x}J_1) = 0 \qquad (10)$$

where the wave-action densities for the pump and backscattered waves are defined as $J_{0,1} \equiv |a_{0,1}|^2 /(2\pi\omega_{0,1})$. With 100 spatial cells, the relative action conservation error was less than 1% in all cases. As a check of the coupled-mode equations, we considered a test case using the linear ion-wave model, a pump-wave intensity below the threshold for absolute instability, $n_e/n_{crit}=0.1$, a Be plasma, $ZT_e/T_i=6.24$, $T_e=2$ keV, $\nu_{ii}/\omega_s=0$, $L_x/\lambda_0 \sim 300$, very weakly seeded backscatter so that pump depletion is negligible, no IAW noise source ($S_{noise}=0$), $\Delta x/L_x=0.01$, $\gamma_{LD}/\omega_s=0.097$, and $v_0/v_e=0.2033$ For this test case, we expect the coupled-mode equations to settle into a steady state with convective growth of the backscatter and a spatial growth rate[1] for the amplitude $a_1$ given by $\kappa_{conv}=\gamma^2_{SBS}/(\gamma_{LD}v_{g1})=5.04\times10^4 m^{-1}$, where $\gamma_{SBS}$ is the homogeneous-medium convective

temporal growth rate and $G^I_{SBBS}=2\kappa_{conv}L_x$ from Eq.(5). This is in good agreement with the coupled-mode-equations integration which yielded a spatial growth rate of $5.0\times10^4 m^{-1}$ after a steady state was reached (~35 ps).

### III. Linear and Nonlinear Ion Wave Features in PIC Simulations

The coupled-mode equations introduced in the preceding section are used to model the more complete particle simulations of SBS. For the results of the coupled-mode equations to agree reasonably well with the kinetic simulations, it is important to use accurate values for the linear and nonlinear ion wave damping and the frequency shifts. To this end we calibrate the ion-wave model in the coupled-mode equations with two types of electrostatic ion wave simulations: initial-value of simulations of undriven, small-amplitude, ion acoustic waves to determine the linear ion wave damping directly (which can be compared to linear theory) and ponderomotively driven ion waves at finite amplitude to motivate our model of nonlinear SBS ion waves.

**A. Ion wave linear damping**

By initializing a small-amplitude, sinusoidal displacement of the ions in a particle simulation to launch a standing wave, a direct measurement of the ion wave frequency and damping rate can be made, which is then compared to linear theory. Figure 1 shows the results from one-dimensional BZOHAR simulations of small-amplitude, ion acoustic, standing waves with $n_e/n_{crit}=0.1$, a Be plasma, $ZT_e/T_i=12$ and 6.24, $T_e=2$ keV, $\nu_{ii}/\omega_s=0$, $L_x/\lambda_{IAW}\sim40$, $k\Delta x=0.04$, $k\lambda_e=0.38$, and 256 and 1024 particles per cell. For $ZT_e/T_i=12$, the BZOHAR simulations agree relatively well with linear theory (including ion thermal effects): $\omega_s=0.269$, $\gamma_s=0.0041$ as compared to $\omega_s=0.28$, $\gamma_s=0.0044$ for $1024/\Delta x$ particles and $\omega_s$=0.272, $\gamma_s$=0.005 for 256/$\Delta$x particles. Note that in BZOHAR's units, the ion plasma frequency $\omega_{pi} = \sqrt{Z/A} = 2/3$ for a Be plasma (Z=4, A=9). For $ZT_e/T_i$=6.24, BZOHAR simulation again agrees relatively well with linear collisionless theory: $\omega_s$=0.293, $\gamma_s$=0.0284 as compared to $\omega_s$=0.31, $\gamma_s$=0.0287 for 1024/$\Delta$x particles (collisionless) and $\omega_s$=0.31, $\gamma_s$=0.029 for 1024/$\Delta$x particles (with collisions, $\nu_0^{NRL}/\omega_s$=0.02). The RPIC[10] particle simulation for the $ZT_e/T_i$=6.24 case yielded $\gamma_s/\omega_s$ =0.097 for 256/$\Delta$x particles using quadratic interpolation for the ion charge density calculation (BZOHAR uses linear interpolation and $\gamma_s/\omega_s$ =0.092 was observed). With fewer particles per cell, the effective collisional diffusion due to discrete particle effects is larger;[11,21] and we expect that the simulations will yield higher damping rates. Ion-ion collisions tend to increase the ion wave damping rates, but there are some subtleties.[25] These test cases provide values for the linear damping rates of the ion acoustic waves needed for the modeling of the SBS simulations in Sec. IV.

**B. BZOHAR simulations of driven ion waves: nonlinear frequency shift and damping with no imposed ion collisions**

We next consider BZOHAR simulations of ponderomotively driven ion waves. We undertook two simulations with values of $ZT_e/T_i$=12 and 6.24. Both simulations exhibit evidence of ion trapping, *e.g.*, an ion velocity tail is produced and there is flattening of the velocity distribution for v~$c_s$. In Figure 2 we show the results of a one-dimensional simulation with $n_e/n_{crit}$=0.1, a Be plasma, $ZT_e/T_i$=12, $T_e$=2 keV, driver turn-on time $\omega_{s0}\tau_{dr}^{on}$=2.5, $\nu_{ii}/\omega_s$=0, $L_x/\lambda_s$~5, $k\Delta x$=0.25, $k\lambda_e$=0.4, driver potential $e\phi_0/T_e$=0.02 and frequency $\Omega=\omega_{s0}=kc_s/(1+k^2\lambda_e^2)^{1/2}$ As a diagnostic we use a Taylor-series expansion of the longitudinal dielectric response with respect to $\varepsilon_{nl}(\omega_{nl},\kappa)=0$ and solve for $\omega_{nl}= \text{Re}\omega_l+\Delta\omega_{nl}+i\gamma_{nl}$ in terms of other quantities using the derived relation:[26,5]

$$k_e^2 \lambda_e^2 \varepsilon(\kappa,\Omega + i\partial/\partial t)\tilde{\phi}(\Omega,\kappa;t) \approx k_e^2 \lambda_e^2 \frac{\partial \varepsilon_{nl}}{\partial \omega}(\Omega - \omega_{nl} + i\partial/\partial t)\tilde{\phi}(\Omega,\kappa;t) = -\tilde{\phi}_0(\Omega,\kappa;t) \quad (11)$$

From the amplitude and phase of the electric potential $\phi$ relative to $\phi_0$, we use Eq.(11) to identify approximate values for the nonlinear frequency shift $\Delta\omega_{nl}$ and damping rate $\gamma_{nl}$. After relaxation of the transients, $\omega_{s0}t/2\pi > 50$, $|e\phi/T_e|\sim 0.1\pm 0.01$, the phase of the electric potential relative to the driving potential approaches $\pi$, $\gamma_{nl}/\omega_{s0} \Rightarrow O(0\pm 0.02)$ and $\Delta\omega_{nl}/\omega_{s0} \Rightarrow -0.12\pm 0.01$ (and both are *not* steady). This can be compared to linear theory, $\gamma_l/\omega_{s0}=0.015$, and nonlinear trapping theory,[13-16] $\Delta\omega_{nl}/\omega_{s0}=-\eta|e\phi/T_e|^{1/2}\sim -0.12$ for $\eta=0.4$ and $\gamma_{nl}\Rightarrow 0$. The full value of the nonlinear frequency shift is realized in the simulation. The damping generally decreases in magnitude in the simulation as time progresses and becomes oscillatory, but the dissipation is positive on average over the simulation.

In Figure 3 we show the results of a one-dimensional BZOHAR simulation with $n_e/n_{\text{crit}}=0.1$, a Be plasma, $ZT_e/T_i=6.24$, $T_e=2$ keV, driver turn-on time $\omega_{s0}\tau_{dr}^{on}=2.5$, $\nu_{ii}^{NRL}/\omega_{s0}=0$, $L_x/\lambda_s\sim 5$, $k\Delta x=0.25$, $k\lambda_e=0.4$, driver potential $e\phi_0/T_e=0.02$ and frequency $\Omega=\omega_{s0}=kc_s/(1+k^2\lambda_e^2)^{1/2}$ Again using Eq.(11), we deduce values for $\Delta\omega_{nl}$ and $\gamma_{nl}$ For $\omega_{s0}t/2\pi > 45$, $|e\phi/T_e|\Rightarrow 0.09$, the phase of the response approaches $\pi$, $\gamma_{nl}/\omega_{s0} \Rightarrow O(0.01\pm 0.02)$ and $\Delta\omega_{nl}/\omega_{s0} \Rightarrow -0.15$ The frequency shift and damping rates are *not* steady, and can be compared to linear theory, $\gamma_l/\omega_{s0}=0.097$, and nonlinear trapping theory,[13-16] $\Delta\omega_{nl}/\omega_{s0}=-\eta|e\phi/T_e|^{1/2}\sim -0.27$ for $\eta=0.9$ and $\gamma_{nl}\Rightarrow 0$. However, the ion wave dissipation remains positive on average. ~60% of the frequency shift expected from nonlinear trapping theory is observed in the simulation.

In Figs. 2 and 3 we also show the results of integrating Eq.(9) from the coupled-mode equations with a fixed ponderomotive driving potential (with a finite turn-on time, $\omega_{s0}\tau_{dr}^{on}\sim 2$) suppressing the dynamics of the electromagnetic waves. In the integration of Eq.(9) we used $\eta=0.4$ and $\eta=0.54$ and $\gamma_{nl}/\omega_{s0}=0.015$ (relaxes to zero) + 0.008 (residual) and $\gamma_{nl}/\omega_{s0}=0.08$ (relaxes to zero) + 0.02 (residual) for $ZT_e/T_i=12$ and 6.24, respectively, as motivated by the BZOHAR simulation results and the frequency-shift and nonlinear-damping diagnostic. The integration of the coupled-mode equations in this case yields qualitatively similar results compared to the BZOHAR results after initial transients relax. Late in time there is some semi-quantitative agreement of the ion-wave magnitudes and the nonlinear frequency shifts. To the extent that the damping rates late in time are small compared to magnitude of the frequency shift and positive, the phases in the kinetic simulation are $\leq \pi$. It is typical that the damping and frequency shift diagnostic based on Eq.(11) yields effective damping rates that are much larger than one would expect based on linear theory early in time, which then decrease to smaller values as transients relax and the velocity distributions flatten due to trapping. Meanwhile the frequency shifts in the BZOHAR simulations deduced from the diagnostic qualitatively resemble those built into the nonlinear ion wave model in the coupled-mode equations. The relative magnitudes of $|\phi/\phi_0|$ and phases in the particle simulations and the couple-mode-equation integrations are similar after the initial transients.

The results of the two simulations for steadily driven ion waves indicate that there is a general relaxation of the ion-wave dissipation and that a nonlinear frequency shift results that approaches the value suggested by trapping theory. However, neither the damping nor the frequency shift is steady; and the time-averaged damping is positive.

The reduced model based on Eq.(9) with a defined ponderomotive driving potential captures some of the important phenomenology at least qualitatively.

**IV. PIC Simulations of SBS and Three-Wave Coupled-Mode Integrations**

In this section we present particle simulations of SBS in one spatial dimension and accompanying integrations of the coupled-mode equations that model the particle simulations. In each of the four studies the incident laser intensity is scanned corresponding to an interesting range of linear gains. As the laser intensity is increased, the SBS reflectivities increase; and nonlinearities become evident.

The rationale for the four test cases is as follows. Firstly, if inflation of SBS occurs in a specific plasma condition, is that occurrence exceptional? To address this question in a limited parameter study, four series of intensity scans were examined. Two values $ZT_e/T_i$=6.24 and 12 are considered, corresponding to relatively strong and weak ion Landau damping. For $ZT_e/T_i$=6.24, we consider one case in which there are imposed ion-ion collisions (Case 1) and another case in which there are no imposed collisions (Case 2). The collisions increase the ion wave damping, which reduces the SBS gain exponent and is expected to reduce the reflectivities. For $ZT_e/T_i$=6.24, we also investigate a case with no imposed ion collisions in which the discrete particle noise and the concomitant fluctuation-driven diffusion have been reduced significantly (Case 3). In Case 3, we expect that trapping can occur at lower ion wave amplitudes because the collisional detrapping is weaker, which circumstance might allow inflation to onset at lower laser intensities. However, the SBS exponentiates from the significantly lower thermal fluctuations for the density perturbations, which leads to a lower saturated reflectivity for the same value of the gain parameter as in Case 2. For $ZT_e/T_i=12$ and no imposed collisions (Case 4), the ion wave linear damping is significantly reduced, which means that the linear gains for SBS are significantly higher for the same values of the laser intensity. The simulations address whether the reduced ion Landau damping in Case 4 makes it easier for inflation to occur.

### A. Case 1: $ZT_e/T_i=6.24$ with ion-ion collisions using BZOHAR

All of the SBS simulations share the following parameters: the laser wavelength corresponds to $\lambda_0=0.35$ μm, $n_e/n_{crit}=0.1$, a Be plasma, $T_e=2$ keV, $L_x/\lambda_0 \sim 300$, and $k_0\Delta x=k_0\lambda_e=0.2$. For this first case $ZT_e/T_i=6.24$, and the ion-ion collision frequency $\nu_{ii}^{NRL}/\omega_{s0}=0.28$, where $\nu_{ii}^{NRL} = 4\pi Z_i^4 e^4 n_i \Lambda_{ii} / m_i^2 v_i^3$, the total relative linear ion wave damping is $\gamma_s^{tot}/\omega_s=0.112$,[25] from the trapping theory $\Delta\omega_{nl}/\omega_{s0}=-0.9(\delta n/n)^{1/2}$, and the simulation durations correspond to $\tau_{sim}\sim 60$ps. The length of the simulation is chosen to approximate a single speckle length, $L_x \sim 8f^2\lambda_0$, where $f=f/\#\sim 6$ is the $f$-number. The ion-ion collisions were imposed using the Fokker-Planck collision model outlined in Ref. 9. For $\delta n/n > 0.015$, which in this series of simulations occurs for $(v_0/v_e)^2>0.02$, the ion response should be in a trapping-dominated physics regime based on Eqs.(1) and (2). The intensity scan $v_0/v_e=0.0707 - 0.2828$ corresponds to linear gains $1 - 16$. Note that $v_0/v_e \approx 6(I_0\lambda_0^2/10^{14}\text{W}\mu\text{m}^2/\text{cm}^2)^{1/2}/T_e(\text{eV})^{1/2}$. Data from the BZOHAR simulations and coupled-mode-equations integrations are shown in Fig. 4. The time-averaged BZOHAR reflectivities follow the Tang formula ($R_{Tang}$) using the linear gains relatively well with a value of $R_0$ so that the resulting reflectivities match the time-averaged BZOHAR reflectivities at the lowest intensities. However, the reflectivities are not steady in the simulations, which motivates the mode-coupling, time-dependent analysis. If the saturated reflectivities were relatively steady and compared well with the Tang formula using the linear ion wave damping rate, then this would be a basis for concluding that there was no significant inflation of SBS.

The BZOHAR simulation peak and time-averaged reflectivities are compared with coupled-mode equation predictions using the linear and nonlinear IAW models in Figs. 4-7. The amplitude $A_{noise}$ of the noise source in Eqs.( 8) and (9) is set by fitting the resulting peak reflectivities to match the BZOHAR peak reflectivities at the lowest intensities for which the SBS signal is above the noise (backscattering off the thermal noise in the density fluctuations). As the value of $v_0/v_e$ increases in Fig. 4a, we see that the BZOHAR peak and time-averaged reflectivities significantly exceed the reflectivities produced by the coupled-mode equations with the linear ion-wave model, which gives evidence of inflation. Use of the nonlinear ion-wave model in the coupled-mode equations yields higher reflectivities and a better match to the BZOHAR data. The "inflation" model ($\Delta\omega_{nl}=0$) reflectivities are much too high.

In Fig. 5 the peak relative electron density perturbations at $x=L_x/4$ in the BZOHAR simulations are compared to couple-mode-equation results as a function of the pump-wave intensity for linear, nonlinear, and "inflation" IAW models. A spatially local comparison is made difficult because the coupled-mode-equation IAW amplitudes are strongly spatially dependent and peak sharply at the left side of the plasma where the pump enters when SBS is strong. Furthermore, the local spatial variability in the unfiltered density perturbation amplitudes is 25-50% throughout. We note that while the coupled-mode equations with the nonlinear model lead to reflectivities that better match the BZOHAR data, use of the linear ion-wave model better matches the ion-wave amplitudes in this particular case.

Figures 6 and 7 show results from a BZOHAR simulation with $v_0/v_e$=0.1414, $ZT_e/T_i$=6.24, $\nu_{ii}^{NRL}/\omega_{s0}$=0.28, $\tau_{sim}$~40ps and the corresponding coupled-mode-equation integrations with linear and nonlinear ion-wave models. Distortion of the ion velocity distribution due to trapping is evident in Fig. 6f. Evidence of coupling of the SBS ion wave to subharmonic features emerges in the electrostatic streak spectrum in Fig. 6e, which may be a manifestation of a secondary instability, *e.g.*, two-ion-wave decay.[5] Mode coupling and two-ion-wave decay processes are omitted from the SBS coupled-mode equations. These processes are a source of nonlinear dissipation for the SBS ion wave, which can reduce the SBS.[5] The mode-coupling results in Fig. 7 capture the bursty time dependence of the BZOHAR reflectivity and the ion wave amplitude in a general sense. We note that the first large burst of SBS reflectivity in Fig. 6a in BZOHAR has the highest reflectivity as is frequently observed in other one and two-dimensional BZOHAR SBS simulations.[6,9] This is not the case in the results from the coupled-mode equations whose bursts of reflectivity do not decrease in magnitude and sometimes increase in time (Fig.7a,c). We have documented that the decrease in the magnitude of the reflectivity bursts in time in the BZOHAR simulations correlates with the occurrence of the coupling/decay of the principal SBS ion wave to lower-frequency, longer-wavelength ion waves.[5,7] We believe that the absence of the coupling of ion waves to other ion waves, including two-ion-wave decay, likely accounts for why the reflectivity bursts are not observed to decrease in magnitude after the first burst in the mode coupling calculations presented here.

**B. Case 2: $ZT_e/T_i=6.24$ with no ion collisions using BZOHAR**

In Fig. 8 we show the results of BZOHAR simulations and corresponding mode-coupling-equation integrations scanning laser intensity with $ZT_e/T_i=6.24$, $T_e=2$ keV, no imposed ion collisions, $\gamma_{LD}/\omega_{s0}=0.092$, $\Delta\omega_{nl}/\omega_{s0}=-0.9(\delta n/n)^{1/2}$, $\tau_{sim}\sim 60$ps. The BZOHAR simulations show ion trapping (Fig. 9), and reflectivities higher than the mode coupling model yield with linear ion waves. In this test case and the two that follow with no imposed ion-ion collisions, evaluation of Eqs.(1) and (2) suggests that the ion wave amplitude thresholds set in for $|\delta n/n|$ values <1% in order to be in the ion-trapping-dominated limit. The intensity scan $v_0/v_e=0.0707 - 0.2828$ corresponds to linear gains 1 – 12. The peak reflectivities and many of the time-averaged reflectivities exceed the Tang formula using the linear gain (Fig. 8a). The peak reflectivities follow the predictions of the coupled-mode equations with the nonlinear ion-wave model (Fig. 8b). Use of the nonlinear and inflation ($\Delta\omega_{nl}/\omega_s=0$) ion-wave models in the coupled-mode equations give higher reflectivities than the reflectivities produced with the linear ion-wave model. The reflectivities using the inflation model are significantly higher than the BZOHAR values (Fig. 8c). Because the BZOHAR peak reflectivities exceed the coupled-mode-equation reflectivities using the linear IAW model, this suggests that there is some inflation of SBS in this intensity scan. In addition, many of the BZOHAR time-averaged reflectivities observed exceed the Tang formula using the linear ion wave damping rate.

Figure 8d shows the mode-coupling reflectivities as functions of laser intensity using a nonlinear ion-wave model amended to include some residual damping ($\Delta\omega_{nlr}\neq 0$, "nlr-c" data), $\text{Im}\omega/\omega_{s0}=0.082$ (relaxes to zero due to trapping)+.01 (residual, does not relax). We note that the collisional discrete particle effects, Eq.(4), motivate the assumption of the residual damping and its approximate magnitude. Furthermore, there is again some evidence of mode coupling and two-ion-wave decay in the BZOHAR electrostatic streak spectrum shown in Fig. 9. The choice of the magnitude of the residual ion wave damping in the mode-coupling model could be made on the basis of a careful assessment of the ion wave damping based on theoretical arguments, or measurements or inferences of the ion wave damping in the kinetic simulations, or by fitting the mode-coupling results (by varying the residual damping) to the results of the kinetic simulations, or some other logic. We did a little exploration of the sensitivity of the mode-coupling results to varying the residual damping, but this exploration was not a systematic optimization with respect to the value of the residual damping. The value of the residual ion wave damping used here should be considered as illustrative and reasonable given Eq.(4).

In Fig. 10 are shown plots of the reflectivities and density perturbations at $x=L_x/4$ as functions of time from coupled-mode integrations with linear and nonlinear (nlr-c) ion-wave models. The reflectivities in the mode-coupling integrations and in BZOHAR are similarly bursty, and the density perturbation amplitudes increase in amplitude with respect to time in both examples in Fig. 10. However, the BZOHAR reflectivity bursts tend to decrease in amplitude with respect to time while the mode-coupling reflectivities increase. We present the peak density perturbation amplitudes at $x=L_x/4$ as functions of $(v_0/v_e)^2$ for BZOHAR and the mode-coupling models in Fig. 11. We note that the mode-coupling data using the nonlinear ion wave model with residual damping (nlr-c) is in better agreement with both the BZOHAR reflectivities (for $(v_0/v_e)^2 > 0.015$) and density perturbations than the data from the other ion-wave models (Figs. 8d and 10). The overall results indicate that the mode-coupling equations augmented with a nonlinear model of the ion waves capture important features in the BZOHAR simulations but do not reproduce everything.

### C. Case 3: $ZT_e/T_i$=6.24 with no ion collisions (RPIC simulation)

We also undertook a series of simulations with the same parameters as in Case 2 using the RPIC particle simulation code,[10,11] which uses quadratic interpolation to compute the ion density and has a lower thermal fluctuation level than does BZOHAR which uses linear interpolation.[11,21] In the RPIC, one-dimensional simulations, the parameters again were $ZT_e/T_i$=6.24, with no imposed ion-ion collisions, $\tau_{sim}$~120-300ps, $\gamma_{LD}/\omega_s$=0.097 (measured independently), and $N_{cell}$=256. The laser intensities were scanned over $v_0/v_e$=0.037 - 0.26 ($6\times10^{13}$W/cm$^2$ - $3\times10^{15}$W/cm$^2$) corresponding to linear gains 1 - 12. With a lower-amplitude thermal noise source, smaller reflectivities and saturated ion wave amplitudes result than were observed in the BZOHAR simulations.

In Fig. 12 we observe that the time-averaged reflectivities follow the Tang formula using the linear gain, but the reflectivities are non-steady. There are a few peak reflectivities in Fig. 12a above the predictions of the mode-coupling model using a linear IAW model (which may suggest that there is some inflation), but the mode-coupling time-averaged reflectivities with a linear ion-wave model agree well with most of the RPIC simulation results. The mode-coupling equations using the nonlinear ion-wave models slightly over-estimate the particle simulation peak and time-averaged reflectivies in Figs. 12b and 12c.

If we include some residual dissipation for the ion waves in the nonlinear ion-wave model (model nlr-c), then the mode-coupling peak and time-averaged reflectivities reproduce the particle simulation reflectivities relatively well using $\gamma_{tot}/\omega_s$=0.082 (relaxes to 0) + 0.015 (residual damping) and retaining a finite nonlinear frequency shift. In fact, both the nlr-c and linear ion wave mode-coupling models fit the particle simulation reflectivities quite well. In Fig. 13 we show an example of the reflectivities as functions of time from a specific particle simulation and in Fig. 14 the corresponding mode-coupling calculations for $I_0$=1.8×10$^{15}$ W/cm$^2$. The mode-coupling calculations in Fig. 14 are similarly bursty, and the magnitudes of the reflectivities are in relatively good agreement. The magnitudes of the reflectivity bursts in the RPIC kinetic simulation and in the mode- coupling integrations here are significantly smaller than those in Case 2 and do not increase in time. The amplitudes of the density perturbations observed in the mode-coupling-equations integrations are similarly smaller here as compared to those in Case 2 for the same laser intensity (and, hence, smaller linear gain exponent) correlating with the reduced thermal noise source in the RPIC simulations out of which the SBS grows. With smaller ion-wave amplitudes, the nonlinear ion-wave effects are weaker. We conclude that in this RPIC series of particle simulations and the corresponding mode-coupling integrations there is almost no evidence of inflation, and the mode-coupling-equations integrations (linear and nlr-c) capture the bursts of reflectivity relatively well.

The power spectrum of the density fluctuation, $|\delta n_e(k,\omega)|^2$, for a representative simulation ($I_0$=1.9×10$^{15}$ W/cm$^2$) is shown in Fig. 15 in two ranges of k and ω for clarity. The top panel shows a non-stimulated noise spectrum, corresponding to the usual linear dispersion of IAWs for $k\lambda_{De}$<0.2, and a stimulated IAW spectral streak at $k\lambda_{De}$~0.38 corresponding to the SBS-generated IAW and its subsequent ion trapping-induced frequency shift. This spectral streak feature is analogous to the Langmuir wave streaks observed in previous RPIC simulations of backward SRS in the (electron) trapping regime.[27] The lower panel of Fig. 15 shows the same power spectrum, but in a smaller range of k and ω. One interesting feature can be seen in this contour plot: the spectral streak is actually composed of several distinct "islands." Each of these "islands" can be correlated to a distinct SBS spatiotemporal pulse moving across the simulation domain.

For each RPIC simulation performed, the density fluctuation at a fixed location (for the cases reported here, this location is 25μm away from the laser entrance boundary) as a funtion of time is recorded. Three significant quantities are calculated from this set of data: the maximum density fluctuation ($\delta n_{max}$), the average of the temporal density maxima ($\delta n_{mean}$), and the root-mean-square (rms) of the temporal density maxima ($\delta n_{rms}$). From the temporal density fluctuation at a fixed location $\delta n_e(x_{fixed},t)$, a set of local maxima of $|\delta n_e(x_{fixed},t)|$ is computed. The largest among these local maxima is then computed and is called, for the lack of better terms, the *maximum density fluctuation* $\delta n_{max}$. $\delta n_{mean}$ and $\delta n_{rms}$ are computed as the *unweighted* arithmetic mean and the rms, respectively, of the set of local maxima. $\delta n_{max}$, $\delta n_{mean}$, and $\delta n_{rms}$ are quantitative measures of the strength of the SBS-generated IAWs. A summary of the results for our RPIC simulations is shown in Fig. 16, along with analogous quantities from three-wave coupled-mode simulations. Similar to the comparison of the BZOHAR ion wave amplitudes to those from the coupled-mode-equations integrations, there is qualitative agreement between the RPIC ion wave amplitudes and those from the coupled-mode-equations modeling. We note that the nonlinear mode-coupling model with residual dissipation (nlr-c) matches both the reflectivities and the trends of the ion wave amplitudes vs. laser intensity relatively well (Figs. 12d and 15). Although there is no indication of inflation in the SBS reflectivities in this RPIC series, the results of the coupled-mode-equation integrations using the nlr-c ion wave model and the significant ion wave amplitudes and evidence of ion trapping effects in the RPIC simulations suggest that ion-wave kinetic nonlinearities are significant elements in these simulations and the plasma response is *not* linear.

**D. Case 4: $ZT_e/T_i=12$ with no ion collisions using bZOHAR**

In this last series of BZOHAR simulations we consider a case with much weaker ion Landau damping and no imposed collisions. The parameters are $ZT_e/T_i=12$, no imposed collisions, $\gamma_s^{LD}/\omega_{s0}=0.015$, $\Delta\omega_{nl}/\omega_{s0}=-0.3(\delta n/n)^{1/2}$, and $\tau_{sim}\sim 35\text{-}80\text{ps}$. We scan laser intensities such that $v_0/v_e=0.04 - 0.1$, which corresponds to linear gains $\sim 2 - 15$. The time-averaged BZOHAR reflectivities mostly follow the Tang formula using the linear gains in Fig. 17, but the reflectivities are non-steady. The peak and time-averaged reflectivities substantially exceed the coupled-mode-equations peak reflectivities using the linear ion-wave model (Fig. 17a), and there is evidence of ion trapping (Fig. 18), suggesting that nonlinearities are inflating the SBS reflectivities.

Integrating the coupled-mode equations with $\Delta\omega_{nl}=0$ ("inflation") and damping that relaxes to zero leads to reflectivities that are systematically too high compared to BZOHAR data (Fig. 17b). The nonlinear IAW models (nlr and nlr-c) that retain nonlinear frequency shifts fit bZohar reflectivity data better than the "linear" and "inflated" models. The nlr-c model includes a small effective collisional damping rate that models discrete particle effects ($\nu_{ii}/\omega_s=0.008$). Use of the nlr-c model leads to a better fit to the bZohar reflectivities and the $\delta n/n$ data in Fig. 18, while the coupled-mode-equation results with the linear IAW model do not fit as well.

Results from a BZOHAR simulation with $v_0/v_e$=0.06, $ZT_e/T_i$=12, $v_0^{NRL}/\omega_s$=0, and $\tau_{sim}$~100ps are shown in Fig. 19. Evidence of mode coupling and two-ion-wave decay again emerges in the electrostatic streak spectrum late in time. The distortion of the velocity distribution due to trapping is clearly evident. The corresponding mode-coupling SBS calculations with the same parameters are shown in Fig. 20 with the linear ion-wave model and with the nonlinear ion-wave model (nlr-c) having finite residual damping $v_{ii}/\omega_s$=0.008 The nonlinear ion model leads to higher IAW amplitudes and reflectivities than with the linear IAW model. The nonlinear ion-wave model with residual damping (nlr-c) compares more favorably with BZOHAR data for both the reflectivities and the ion wave density perturbations in Figs. 17-20.

## V. Conclusions

Four series of particle simulations and accompanying coupled-mode-equation integrations were undertaken in which the laser intensity was scanned looking for evidence of SBS inflation due to nonlinear effects. For the cases with $ZT_e/T_i$=6.24 with and without imposed collisions using BZOHAR (Cases 1 and 2), we saw significant evidence of inflation and generally similar behavior, although Case 2 had larger linear gains and slightly higher saturated reflectivities. The ion-wave peak amplitudes in Case 1 are smaller than in the collisionless case, Case 2. In Case 3 ($ZT_e/T_i$=6.24 with no imposed collisions, using the RPIC simulation code), the significantly lower thermal fluctuations due to discrete-particle noise led to a factor of 10 lower thermal Thomson backscattering noise signal in the reflectivities than in Case 2, and approximately a factor of 10 lower reflectivities overall. With $ZT_e/T_i$=12 in Case 4 using BZOHAR with no imposed collisions, the results are qualitatively similar to Case 2 with $ZT_e/T_i$=6.24, in the magnitude of the inflation effects, the significant reflectivities and the ion trapping effects observed, the evidence of mode coupling and/or two-ion-wave decay, and the general importance of nonlinearities.

Over the range of parameters examined in three of the four intensity scans using our particle simulations of SBS there is evidence of inflation, *i.e.*, either the peak or both the peak and time-averaged reflectivities in the particle simulations exceed the reflectivities resulting from the corresponding coupled-mode-equation integrations with a linear IAW model for some range of pump-wave intensities. The amount of inflation in the reflectivities varies from a factor of a few to a factor of ten. In the Case 3 intensity scan the SBS grows from a smaller thermal fluctuation level leading to lower reflectivities and lower saturated ion-wave amplitudes with weaker nonlinear effects and little or no evidence of inflation.

It is interesting that the predictions of the Tang formula using the linear ion wave damping rate compare relatively well with the time-averaged reflectivities in the particle simulations. Because the reflectivities are quite time dependent in these simulations, the Tang formula is not really applicable. Furthermore, the Tang formula ignores all nonlinearities in the ion waves and in the ion velocity distribution perturbations, although it does capture the important pump-depletion nonlinearity. Perhaps the success of the Tang formula in predicting the time-averaged reflectivities observed in the particle simulations is only fortuitous.

The coupled-mode-equation integrations with linear and nonlinear models for the ion waves have been used to compare to the particle simulation results. The nonlinear ion-wave models embody the basic phenomenology of ion trapping which can nonlinearly reduce the ion Landau damping and produce nonlinear frequency shifts. Because the value of the ion wave damping is a very influential parameter in the modeling, we have undertaken direct particle simulations of ion waves to measure the ion wave damping in the small-amplitude, linear-physics regime. We have also performed simulations of ponderomotively driven ion waves in electrostatic simulations to assess finite-amplitude effects on the ion wave damping and to measure frequency shifts directly. Our modeling of the particle simulations with the coupled-mode equations indicates that use of a linear ion-wave model is inadequate for most of the simulation results when both the reflectivities and the ion-wave density perturbations exceed a few per cent. In three of the four intensity scans (Test Cases 1, 2, and 4), use of the nonlinear ion-wave models ("nlr" in Case 1 and "nlr-c" in Cases 2 and 4) in the mode coupling equations leads to results that better match the particle simulation results. In Test Case 3, there is little difference in the results of the coupled-mode equations using the linear and "nlr-c" models, both of which lead to reflectivities that match both the peak and time-averaged reflectivities well. However, use of the "nlr-c" model agrees better with the observed ion wave amplitudes than does using the linear ion wave model in Case 3. The agreement of the coupled-mode equations with the particle simulation reflectivities in Cases 2 and 4 also was improved with additions of small amounts of residual dissipation or effective collisionality ("nlr-c" model, which takes into account the discrete-particle effects in the PIC simulations or other sources of residual IAW damping). In all of the cases studied, use of a nonlinear ion-wave model that completely suppresses the nonlinear frequency shift leads to overestimates of the reflectivities and the ion wave amplitudes.

The experience here with the reduced modeling of SBS using coupled-mode equations augmented with a nonlinear model of the ion waves is sufficiently encouraging to motivate further work with the coupled-mode equations in which we extend and improve the coupled-mode-equations model as guided by kinetic particle simulations. The extensions address certain physics omissions. We have begun incorporating a model of ion-wave mode coupling and/or two-ion-wave decay in the SBS coupled-mode-equations model, because we believe that ion wave mode coupling/two-ion-wave decay is a potentially important ion wave dissipation mechanism. Another omission in our mode-coupling model is that the ion trapping flattens and distorts the ion velocity distribution function, which distortions survive when the ion wave relaxes unless there are ion side losses or collisional relaxation. The distortion of the ion velocity distribution alters the damping rate of the ion wave and its frequency.[7,19] A reduced model of the ion trapping effects on SBS that includes the evolution of the flattening of the ion velocity distribution by introducing equations for the growth and relaxation of the ion velocity plateau width has been introduced in Ref. 19. We will compare the results of the coupled-mode-equations model to kinetic SBS simulations incorporating these extensions in the future.

The one-dimensional particle simulations of SBS and the modeling with coupled-mode equations presented here suggest that inflation can occur. Our previous work in two-dimensional SBS simulations[6,9] has indicated that ion trapping effects are just as vigorous and the effects of ion wave mode coupling and two-ion-wave decay are profound in saturating SBS in two spatial dimensions. Whether the inflation observed here in one dimension survives in two dimensions remains to be investigated. We note that we are unaware of any experimental evidence for significant kinetic inflation of SBS.

**Acknowledgements**

This work was performed under the auspices of the U.S. Dept. of Energy by the University of California Lawrence Livermore National Laboratory under contract No. W-7405-ENG-48. H. X. Vu was supported by the NNSA under the Stewardship Science Academic Alliances Program through DOE Research Grant # DE-FG52-04NA00141/A000. We thank Don Dubois, Laurent Divol, Bruce Langdon for their interest and encouragement, Bedros Afeyan for suggesting comparing coupled-mode equation integrations to the particle simulation results, and Dick Berger for useful input, suggestions, and a careful reading of the manuscript.

Figure Captions

FIG. 1. $|e\phi/T_e|^2$ vs. time in initial-value BZOHAR simulations of undriven small-amplitude ion waves for $ZT_e/T_i=12$, with (a) 1024 particles per cell and (b) and 256 particles per cell, and for $ZT_e/T_i=6.24$, with 1024 particles per cell and (c) with no collisions or (d) with collisions, $\nu_0^{NRL}/\omega_s=0.02$

FIG. 2. BZOHAR electrostatic simulation and corresponding integration of the coupled-mode equation Eq.(9) for a ponderomotively driven ion wave with $ZT_e/T_i=12$ and driver amplitude $e\phi_0/T_e = 0.02$ (a) $|\tilde{\phi}/\tilde{\phi}_0|$ vs. time, (b) phase of $|\tilde{\phi}/\tilde{\phi}_0|$ vs. time, (c) $\Delta\omega_{nlr}/\omega_{s0}$ vs. time, and (d) $\gamma_{nlr}/\omega_{s0}$ vs. time, where $\tilde{\phi}$ and $\tilde{\phi}_0$ are amplitudes of the envelopes for the self-consistent electric potential (determined by Poisson's equation) and the imposed ponderomotive potential.

FIG. 3. BZOHAR electrostatic simulation and corresponding integration of the coupled-mode equation Eq.(9) for a ponderomotively driven ion wave with $ZT_e/T_i=6.24$ and driver amplitude $e\phi_0/T_e = 0.02$ (a) $|\tilde{\phi}/\tilde{\phi}_0|$ vs. time, (b) phase of $|\tilde{\phi}/\tilde{\phi}_0|$ vs. time, (c) $\Delta\omega_{nlr}/\omega_{s0}$ vs. time, and (d) $\gamma_{nlr}/\omega_{s0}$ vs. time.

FIG. 4. Peak and time-averaged reflectivities as functions of $|v_0/v_e|^2$ from SBS BZOHAR particle simulations with $ZT_e/T_i=6.24$ and imposed Fokker-Planck ion-ion collisions (with ion wave dissipation due to collisions $\nu_{ii}/\omega_{s0} = 0.02$) compared to mode-coupling-equations integrations with (a) linear ion-wave model, (b) inflation ion-wave model ($\Delta\omega_{nlr} = 0$), and (c) nonlinear ion-wave model. Also shown is the steady-state reflectivity determined by the Tang formula, Eq.(6). (color on-line)

FIG. 5. Peak ion-wave, relative density perturbation $|\delta n_e/n_e|$ at $x=L_x/4$ as function of $|v_0/v_e|^2$ from SBS BZOHAR particle simulations with $ZT_e/T_i=6.24$ and imposed Fokker-Planck ion-ion collisions compared to mode-coupling-equations integrations. (color on-line)

FIG. 6. From a BZOHAR SBS simulation with $ZT_e/T_i=6.24$, imposed Fokker-Planck ion-ion collisions, and $|v_0/v_e|=0.1414$ (a) instantaneous and cumulative time-averaged reflectivities, $R$ and $<R>$, respectively, vs. time; (b) $e\phi/T_e$ vs. time at $x=L_x/4$; (c) $-n_e/n_c$ vs. time at $x=L_x/4$; (d) streak spectrum for the reflected electromagnetic power at the pump-laser incident plane, power vs. frequency shift from the incident pump wave in units of $\omega_s$ and time; (e) streak spectrum $|e\phi/T_e|^2$ at $x=L_x/4$ vs. frequency in units of $\omega_s$ and time; (f) longitudinal ion velocity distribution function $f(u_x)$ vs. $u_x$. (color on-line)

FIG. 7. From SBS mode-coupling-equation integrations $ZT_e/T_i=6.24$, $\nu_{ii}/\omega_{s0}=0.02$, and $|v_0/v_e|=0.1414$ using a linear ion-wave model, (a) instantaneous and cumulative time-averaged reflectivities, $R$ and $<R>$, respectively, vs. time; (b) $|\delta n_e/n_e|$ at $x=L_x/4$ vs. time; and using a nonlinear ion-wave model including collisional damping, (c) instantaneous and cumulative time-averaged reflectivities, $R$ and $<R>$, respectively, vs. time; (d) $|\delta n_e/n_e|$ at $x=L_x/4$ vs. time. (color on-line)

FIG. 8. Peak and time averaged reflectivities as functions of $|v_0/v_e|^2$ from SBS BZOHAR particle simulations with $ZT_e/T_i=6.24$ and no collisions compared to mode-coupling-equations integrations with (a) linear ion-wave model, (b) nonlinear ion-wave model, (c) inflation ion-wave model, and (d) nonlinear-with-effective-collisions ion-wave model.

Also shown is the steady-state reflectivity determined by the Tang formula, Eq.(6). (color on-line)

FIG. 9. From a BZOHAR SBS simulation with $ZT_e/T_i=6.24$, no collisions, and $|v_0/v_e|=0.13$ (a) instantaneous and cumulative time-averaged reflectivities, $R$ and $<R>$, respectively, vs. time; (b) $e\phi/T_e$ vs. time at $x=L_x/4$; (c) $-n_e/n_c$ vs. time at $x=L_x/4$; (d) streak spectrum for the reflected electromagnetic power at the pump-laser incident plane, power vs. frequency shift relative to the incident pump wave in units of $\omega_s$ and time; (e) streak spectrum $|e\phi/T_e|^2$ at $x=L_x/4$ vs. frequency in units of $\omega_s$ and time; (f) longitudinal ion velocity distribution function $f(u_x)$ vs. $u_x$. (color on-line)

FIG. 10. From SBS mode-coupling-equation integrations with $ZT_e/T_i=6.24$, nominally collisionless, residual damping in the nlr-c model $v_{ii}/\omega_{s0}=0.01$, and $|v_0/v_e|=0.13$ using a linear ion-wave model, (a) instantaneous and cumulative time-averaged reflectivities, $R$ and $<R>$, respectively, vs. time; (b) $|\delta n_e/n_e|$ at $x=L_x/4$ vs. time; and using a nonlinear ion-wave model including collisional damping, (c) instantaneous and cumulative time-averaged reflectivities, $R$ and $<R>$, respectively, vs. time; (d) $|\delta n_e/n_e|$ at $x=L_x/4$ vs. time. (color on-line)

FIG. 11. Peak ion-wave, relative density perturbation $|\delta n_e/n_e|$ at $x=L_x/4$ as function of $|v_0/v_e|^2$ from SBS BZOHAR particle simulations with $ZT_e/T_i=6.24$ and no imposed ion collisions compared to mode-coupling-equations integrations. (color on-line)

FIG. 12. Peak and time-averaged reflectivities as functions of $|v_0/v_e|^2$ from SBS using the RPIC particle simulation code with $ZT_e/T_i=6.24$ and no collisions compared to mode-coupling-equations integrations with (a) linear ion-wave model, (b) nonlinear ion-wave

model, (c) inflation ion-wave model, and (d) nonlinear-with-effective-collisions ion-wave model with $v_{ii}/\omega_{s0} = 0.015$. Also shown is the steady-state reflectivity determined by the Tang formula, Eq.(6). (color on-line)

FIG.13. From a SBS simulation using the RPIC particle simulation code with $ZT_e/T_i=6.24$, no collisions, and laser intensity $1.8\times10^{15}$ W/cm$^2$ (a) instantaneous reflectivity $R$ and (b) cumulative time-averaged reflectivity $<R>$ vs. time.

FIG. 14. From a SBS simulation using the coupled-mode equations with $ZT_e/T_i=6.24$, no collisions, and laser intensity $1.8\times10^{15}$ W/cm$^2$, and a linear ion-wave model, (a) instantaneous and cumulative time-averaged reflectivities, $R$ and $<R>$, vs. time, (b) $|\delta n_e/n_e|$ at $x=L_x/4$ vs. time, and using a nonlinear ion-wave model with residual damping $v_{ii}/\omega_{s0} = 0.015$, (c) instantaneous and cumulative time-averaged reflectivities, $R$ and $<R>$, vs. time, (d) $|\delta n_e/n_e|$ at $x=L_x/4$ vs. time. (color on-line)

FIG. 15. Ion wave power spectrum as a function of wavenumber $k$ and frequency $\omega$ computed over the entire time interval of an RPIC simulation for $ZT_e/T_i=6.24$ and $I_0=1.9\times10^{15}$ W/cm$^2$.

FIG. 16. Ion wave amplitudes at $x=L_x/4$ : the temporal maximum $|\delta n_e/n_e|_{max}$ and the mean over time of the local maxima in an RPIC simulation $\langle|\delta n_e/n_e|_{max}\rangle$; and the mode-coupling-equation results for the maximum of $|\delta n_e/n_e|$ for linear and nlr-c ion wave models vs. $|v_0/v_e|^2$ for $ZT_e/T_i=6.24$ and no collisions. (color on-line)

FIG. 17. Peak and time-averaged reflectivities as functions of $|v_0/v_e|^2$ from SBS using the BZOHAR particle simulation code with $ZT_e/T_i=12$ and no collisions compared to mode-coupling-equations integrations with (a) linear ion-wave model, (b) nonlinear ion-wave

model, (c) inflation ion-wave model, and (d) nonlinear-with-effective-collisions ion-wave model with $v_{ii}/\omega_{s0} = 0.008$. Also shown is the steady-state reflectivity determined by the Tang formula, Eq.(6). (color on-line)

FIG. 18. Peak ion-wave relative density perturbation $|\delta n_e/n_e|$ at $x=L_x/4$ as function of $|v_0/v_e|^2$ from SBS BZOHAR particle simulations with $ZT_e/T_i=12$ and no imposed ion collisions compared to mode-coupling-equations integrations. (color on-line)

FIG. 19. From a BZOHAR SBS simulation with $ZT_e/T_i=12$, no collisions, and $|v_0/v_e|=0.06$ (a) instantaneous and cumulative time-averaged reflectivities, $R$ and $<R>$, respectively, vs. time; (b) $e\phi/T_e$ vs. time at $x=L_x/4$; (c) $-n_e/n_c$ vs. time at $x=L_x/4$; (d) streak spectrum for the reflected electromagnetic power at the pump-laser incident plane, power vs. frequency shift from the incident pump wave in units of $\omega_{s0}$ and time; (e) streak spectrum $|e\phi/T_e|^2$ at $x=L_x/4$ vs. frequency in units of $\omega_{s0}$ and time; (f) longitudinal ion velocity distribution function $f(u_x)$ vs. $u_x$. (color on-line)

FIG. 20. From a SBS simulation using the coupled-mode equations with $ZT_e/T_i=12$, no collisions, and $|v_0/v_e|=0.06$, and a linear ion-wave model, (a) instantaneous and cumulative time-averaged reflectivities, $R$ and $<R>$, vs. time, (b) $|\delta n_e/n_e|$ at $x=L_x/4$ vs. time, (c) and $|\delta n_e/n_e|$ vs. $x=L_x/4$ at $t=120$ps, and using a nonlinear ion-wave model with residual damping $v_{ii}/\omega_{s0} = 0.008$, (d) instantaneous and cumulative time-averaged reflectivities, $R$ and $<R>$, vs. time, (e) $|\delta n_e/n_e|$ at $x=L_x/4$ vs. time, (f) $|\delta n_e/n_e|$ vs. $x=L_x/4$ at $t=120$ps (color on-line)

Figure 1

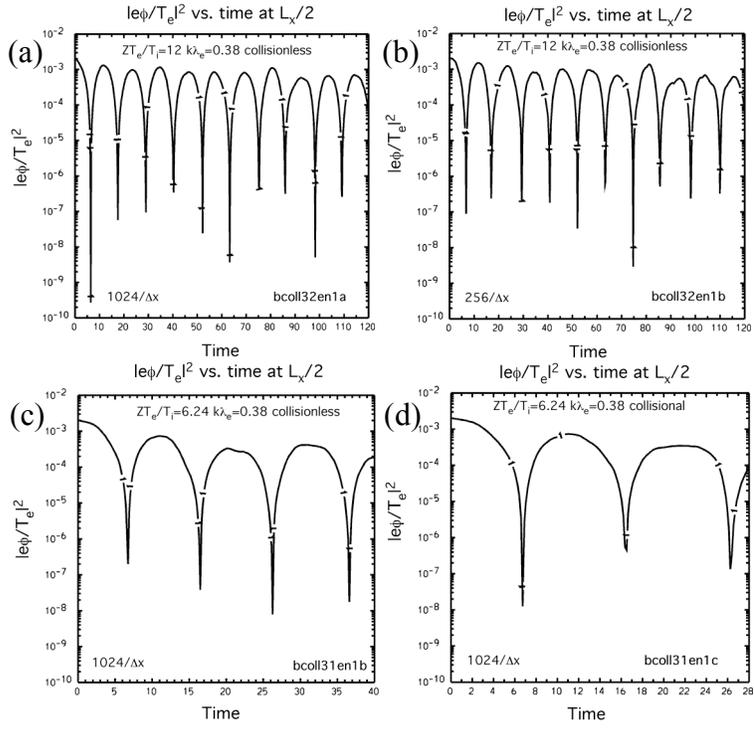

Figure 2

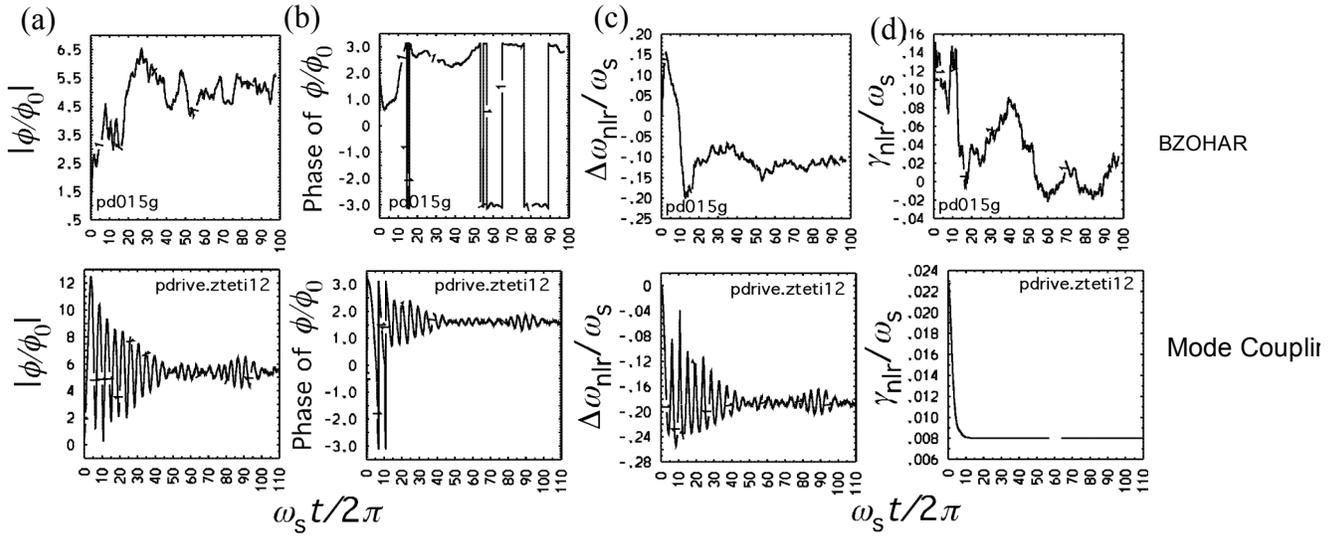

Figure 3

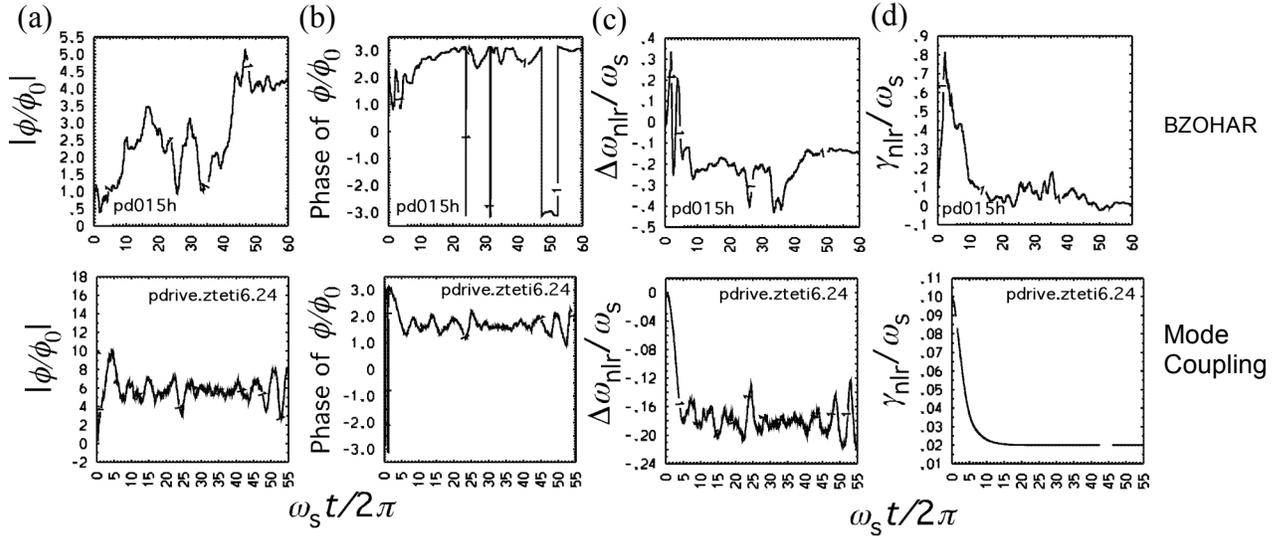

Figure 4

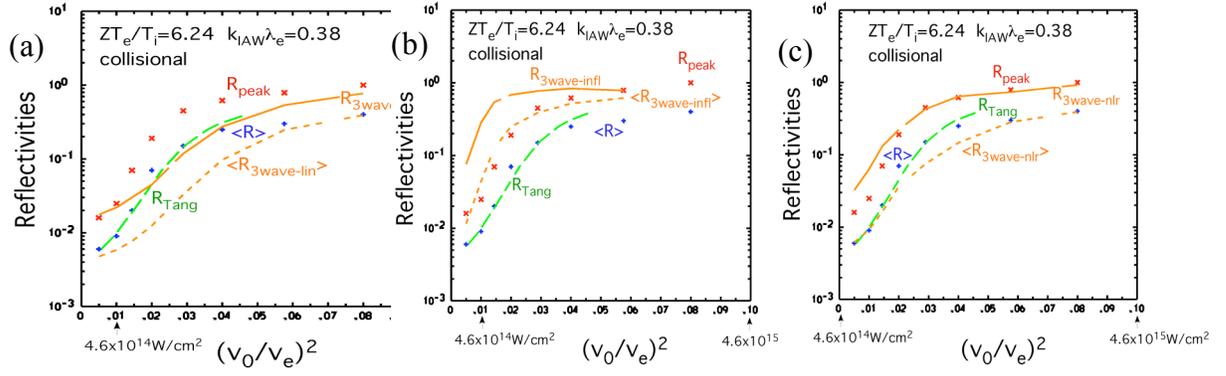

Figure 5

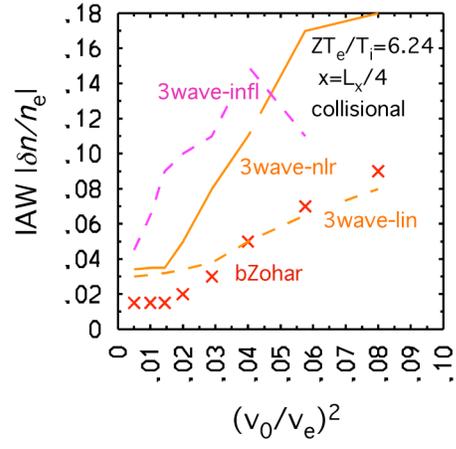

Figure 6

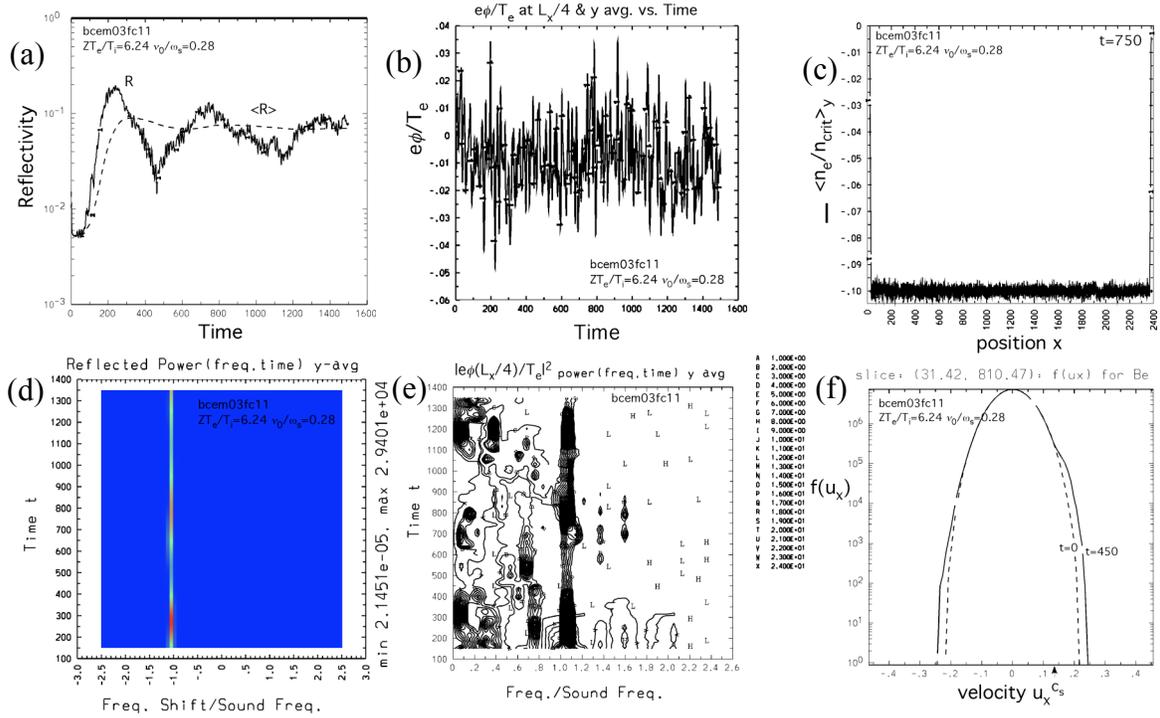

Figure 7

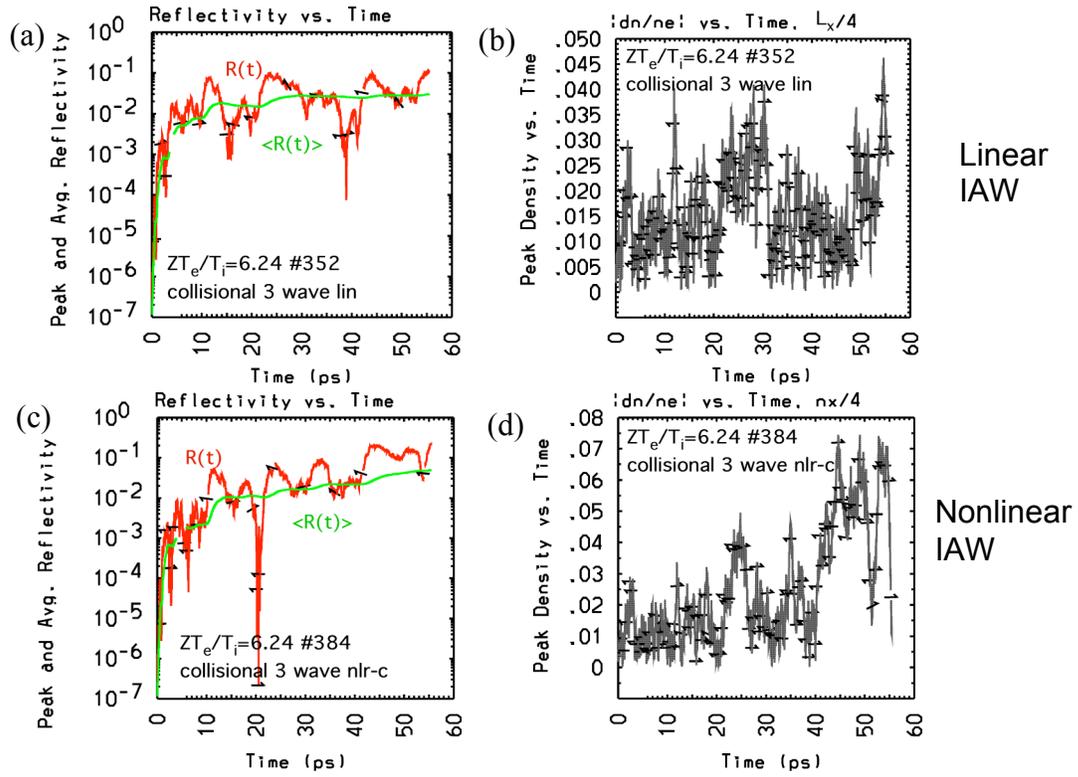

Figure 8

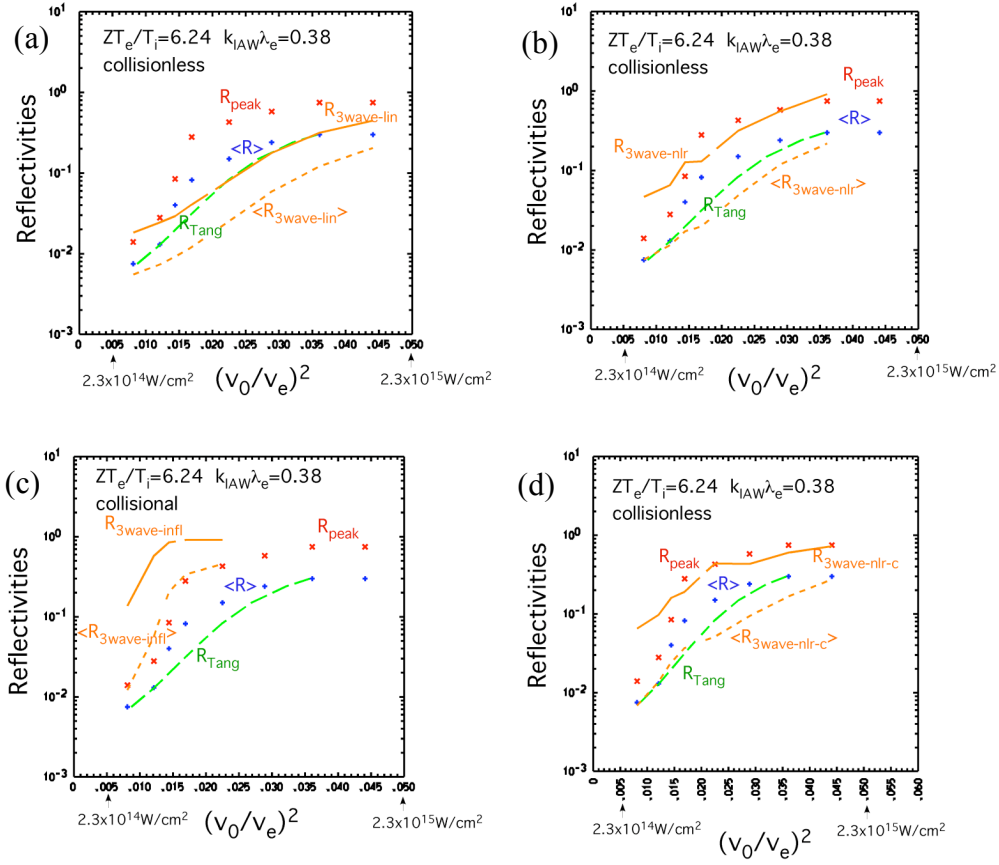

Figure 9

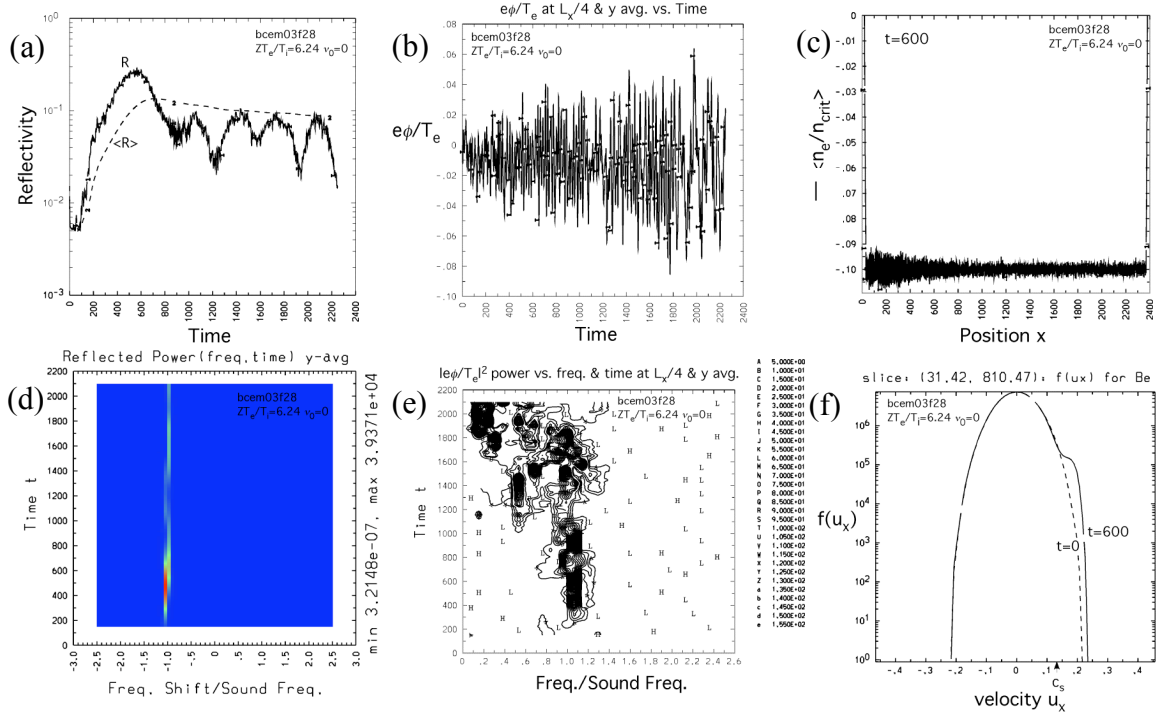

Figure 10

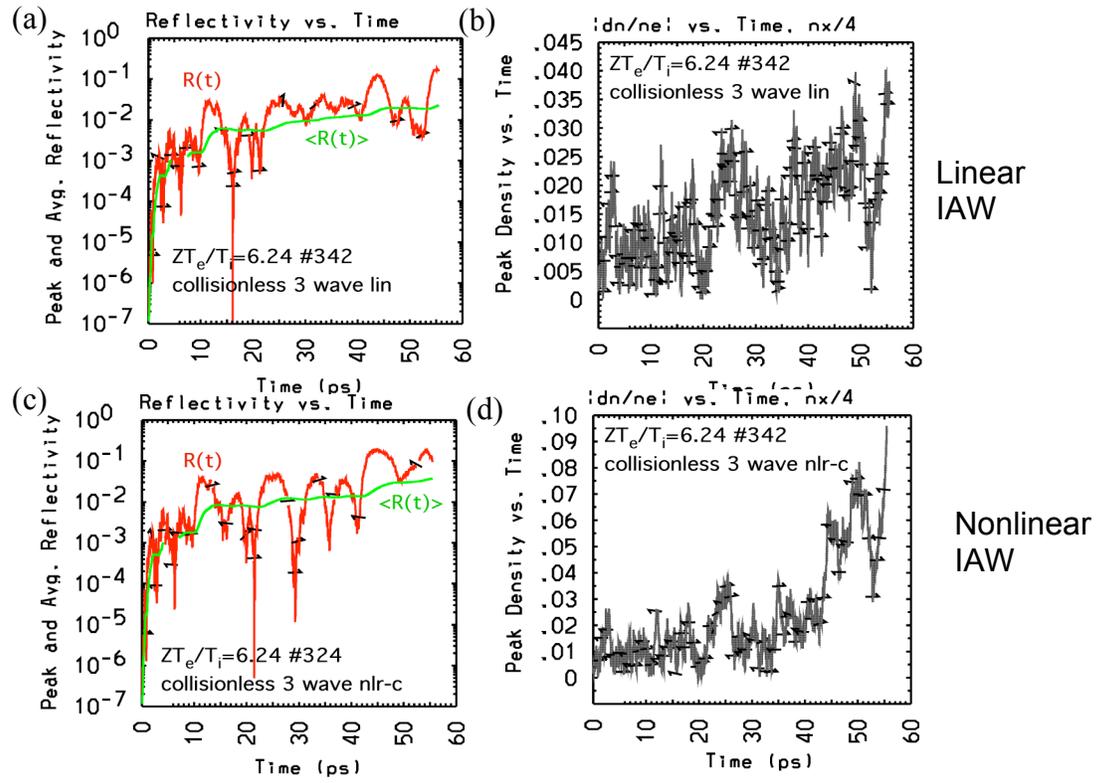

Linear IAW

Nonlinear IAW

Figure 11

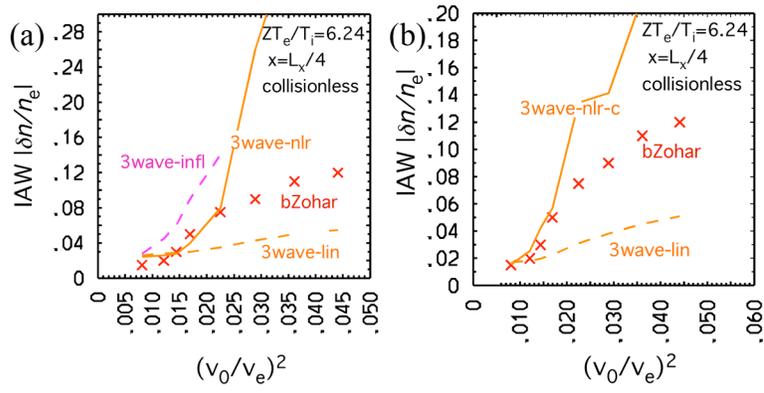

Figure 12

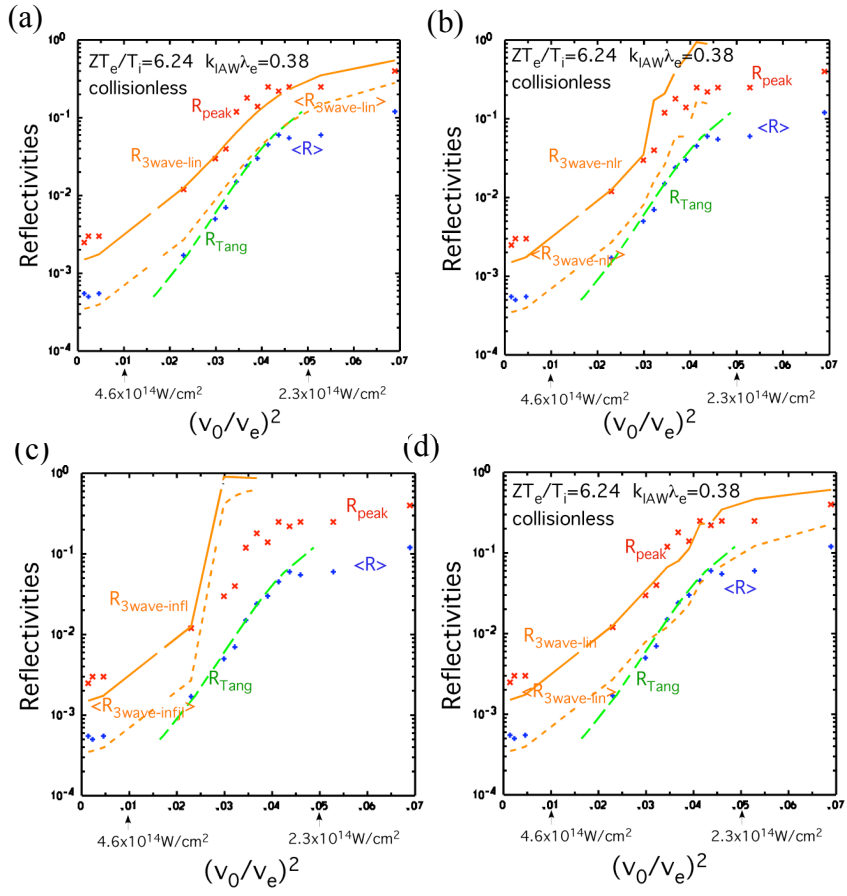

Figure 13

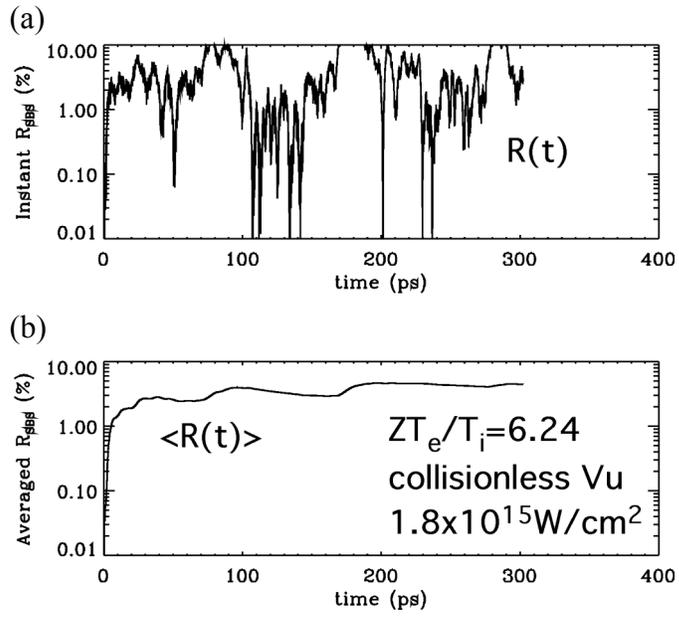

Figure 14

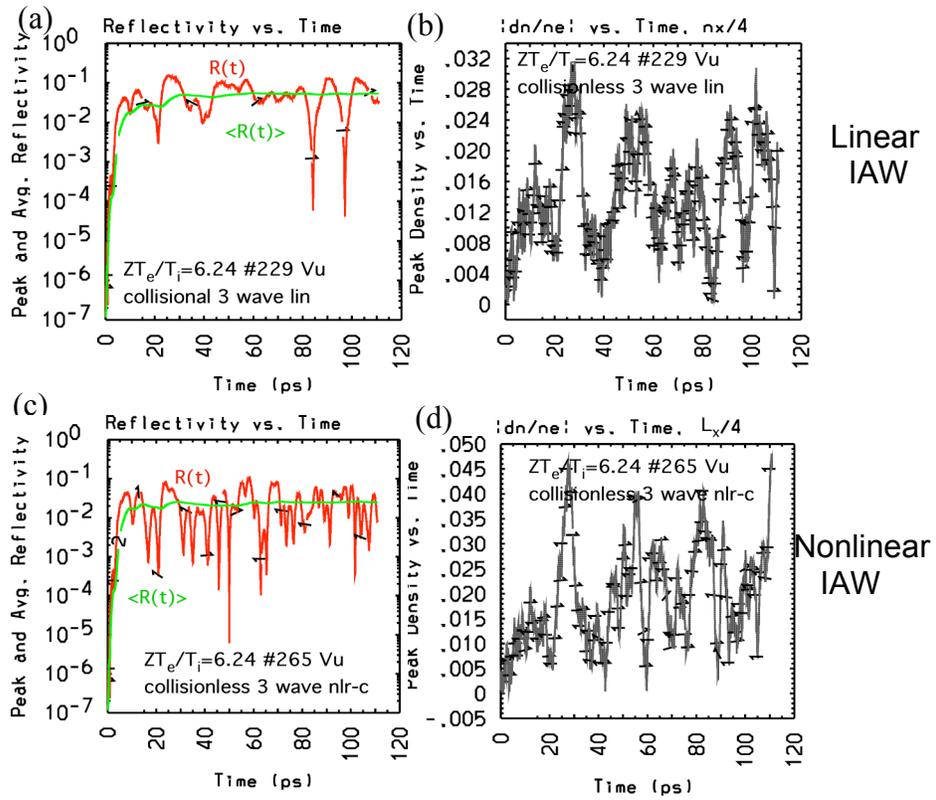

Figure 15

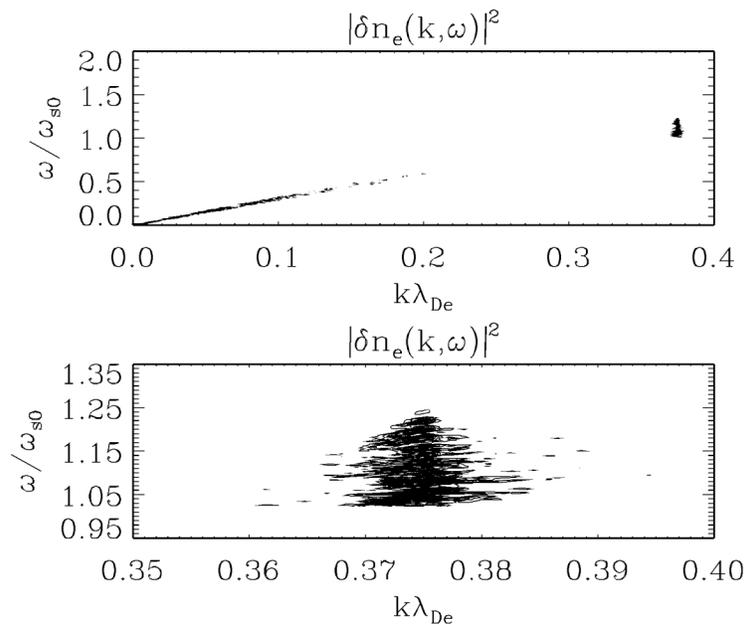

Figure 16

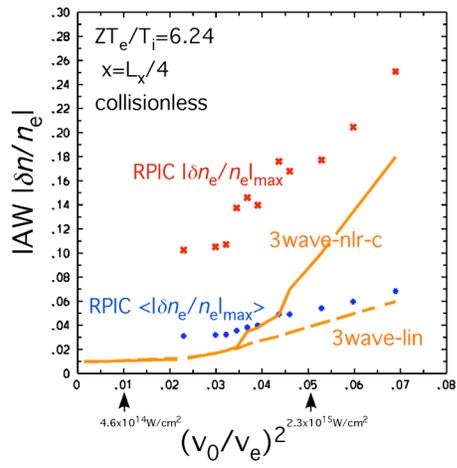

Figure 17

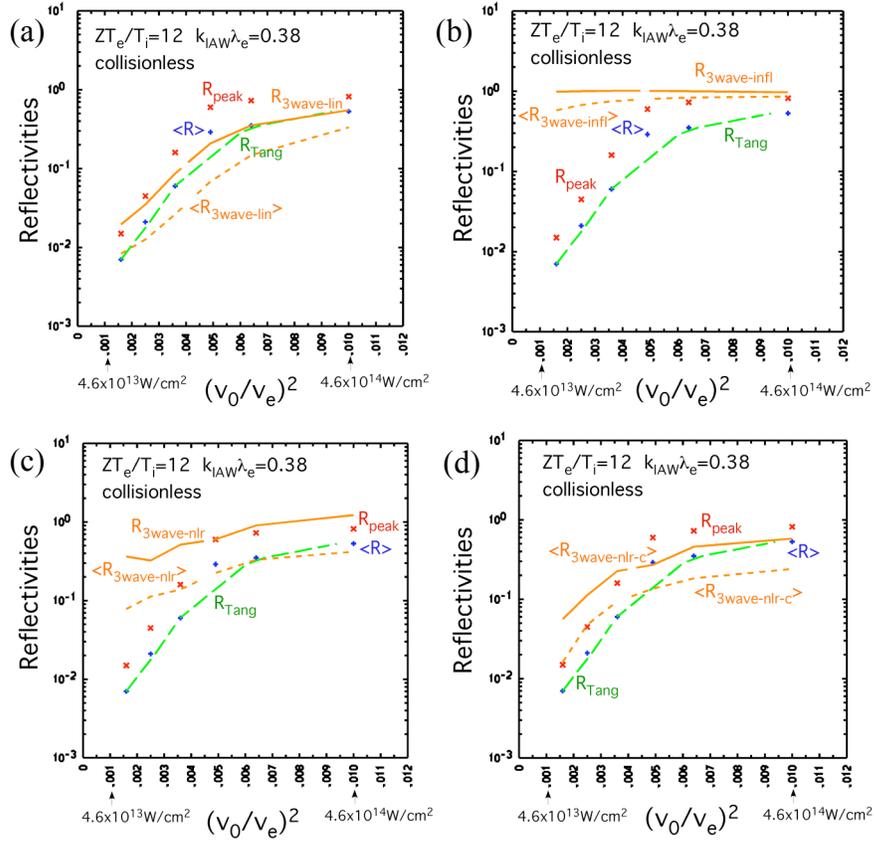

Figure 18

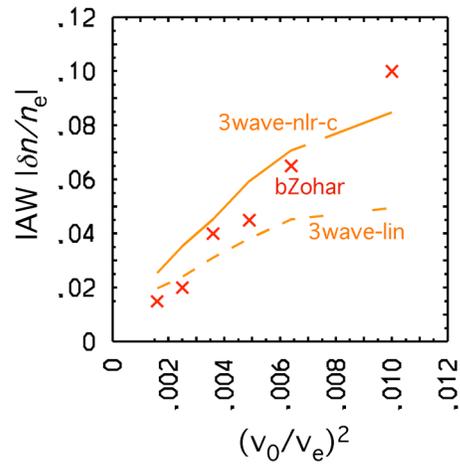

Figure 19

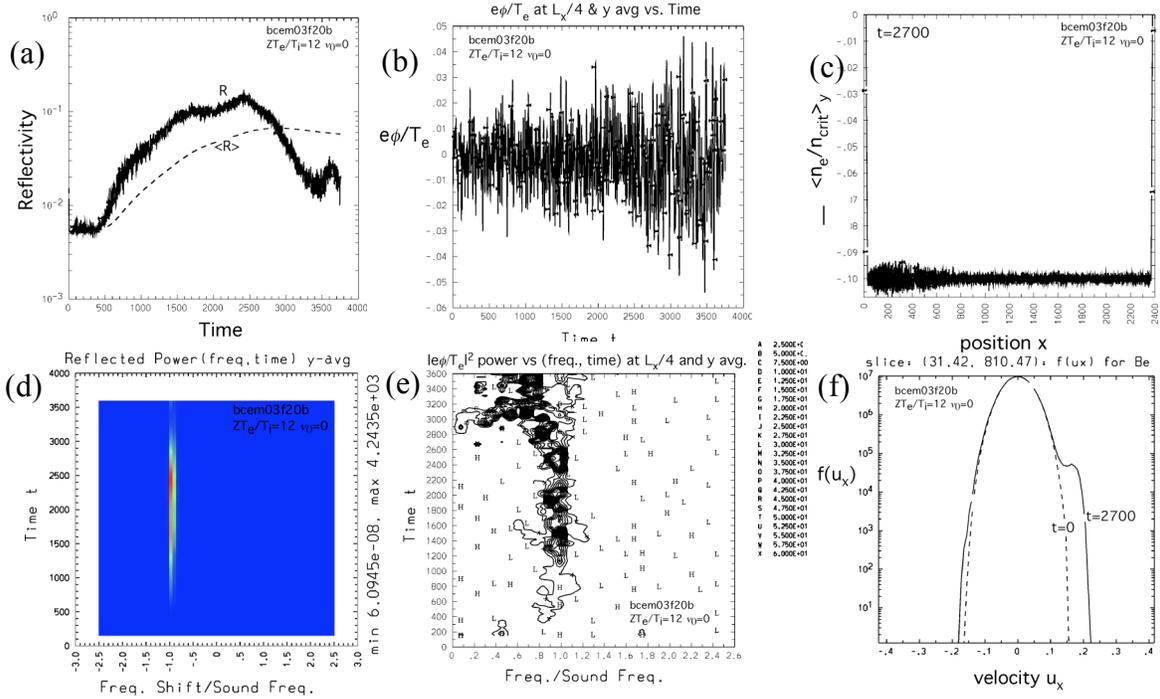

Figure 20

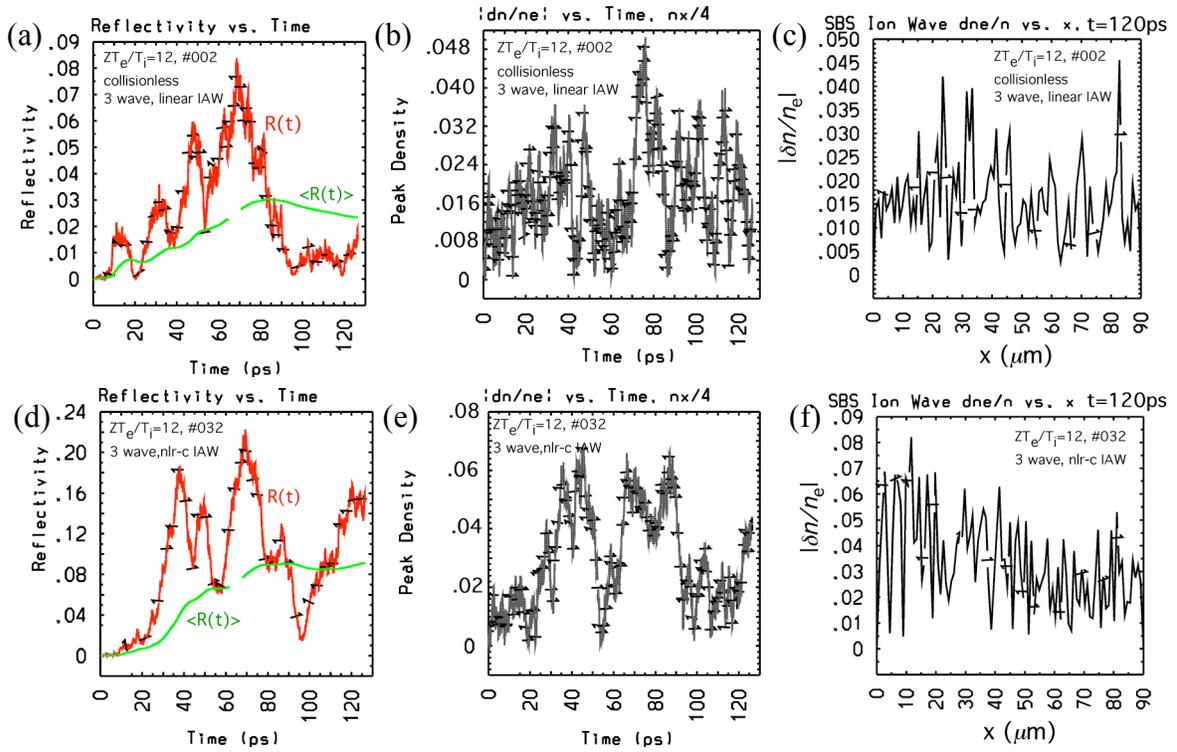